\documentclass[preprint]{aastex63}
\usepackage[utf8]{inputenc}

\usepackage{amsmath,amssymb,amstext}
\usepackage{booktabs}

\usepackage{epstopdf}

\usepackage{amsmath,amssymb,amstext}

\usepackage{natbib}
\usepackage{color}
\definecolor{darkred}{RGB}{175,0,0}

\makeatletter

\newcommand{\Rmnum}[1]{\expandafter\@slowromancap\romannumeral #1@}
\makeatother

\begin{document}

\title{MEM$\_$GE: a new maximum entropy method for image reconstruction \\ from solar X-ray visibilities}

\author{Paolo Massa}
\affil{Dipartimento di Matematica, Universit\`a di Genova, via Dodecaneso 35 16146 Genova, Italy}
\email{massa.p@dima.unige.it}

\author{Richard Schwartz}
\affil{NASA Goddard Space Flight Center, Greenbelt (MD), USA}
\email{richard.a.schwartz@nasa.gov}

\author{A Kim Tolbert}
\affil{NASA Goddard Space Flight Center, Greenbelt (MD), USA}
\email{anne.k.tolbert@nasa.gov}

\author{Anna Maria Massone}
\affil{Dipartimento di Matematica, Universit\`a di Genova, via Dodecaneso 35 16146 Genova, Italy}
\affil{CNR - SPIN Genova, via Dodecaneso 33 16146 Genova, Italy}
\email{massone@dima.unige.it}

\author{Brian R Dennis}
\affil{NASA Goddard Space Flight Center, Greenbelt (MD), USA}
\email{brian.r.dennis@nasa.gov}

\author{Michele Piana}
\affil{Dipartimento di Matematica, Universit\`a di Genova, via Dodecaneso 35 16146 Genova, Italy}
\affil{CNR - SPIN Genova, via Dodecaneso 33 16146 Genova, Italy}
\email{piana@dima.unige.it}

\author{Federico Benvenuto}
\affil{Dipartimento di Matematica, Universit\`a di Genova, via Dodecaneso 35 16146 Genova, Italy}
\email{benvenuto@dima.unige.it}

%\title{MEM$\_$GE: a new maximum entropy algorithm for image reconstruction from {\em{RHESSI}} and {\em{STIX}} visibilities}
%\author{Paolo Massa$^1$, Michele Piana$^{1,2}$, Anna Maria Massone$^{1,2}$ and Federico Benvenuto$^{1}$}
%\date{September 2019}
%\address{$^1$Dipartimento di Matematica, Universit\`a di Genova, Via Dodecaneso 35, 16146, Genova, Italy\\
%$^2$CNR - SPIN, Via Dodecaneso 33, 16146, Genova, Italy}

%\maketitle

\begin{abstract}
    Maximum Entropy is an image reconstruction method conceived to image a sparsely occupied field of view and therefore particularly appropriate to achieve super-resolution effects. Although widely used in image deconvolution, this method has been formulated in radio astronomy for the analysis of observations in the spatial frequency domain, and an Interactive Data Language (IDL) code has been implemented for image reconstruction from solar X-ray Fourier data. However, this code relies on a non-convex formulation of the constrained optimization problem addressed by the Maximum Entropy approach and this sometimes results in unreliable reconstructions characterized by unphysical shrinking effects. 
    
    This paper introduces a new approach to Maximum Entropy based on the constrained minimization of a convex functional. In the case of observations recorded by the {\em{Reuven Ramaty High Energy Solar Spectroscopic Imager (RHESSI)}}, the resulting code provides the same super-resolution effects of the previous algorithm, while working properly also when that code produces unphysical reconstructions. Results are also provided of testing the algorithm with synthetic data simulating observations of the {\em{Spectrometer/Telescope for Imaging X-rays (STIX)}} in {\em{Solar Orbiter}}. The new code is available in the  {\em{HESSI}} folder of the {\em{Solar SoftWare (SSW)}} tree.
\end{abstract}


\section{Introduction}
Solar SoftWare (SSW) contains a large collection of computational procedures for the reconstruction of images from the X-ray data recorded by the {\em{Reuven Ramaty High Energy Solar Spectroscopic Imager (RHESSI)}} \citep{2002SoPh..210....3L} in the time interval between February 2002 to August 2018 - see \citet{2019ApJ...887..131D} for a recent evaluation of the performance of the different available methods. Some of these methods apply directly to {\em{RHESSI}} counts while others have been conceived to process {\em{RHESSI}} visibilities, i.e. calibrated samples of the Fourier transform of the incoming photon flux, generated via a data stacking process. Among count-based methods, SSW includes Back Projection \citep{2002SoPh..210...61H}, Clean \citep{1974A&AS...15..417H}, Forward Fit \citep{2002SoPh..210..193A}, Pixon \citep{1996ApJ...466..585M}, and Expectation Maximization \citep{2013A&A...555A..61B}; among visibility-based methods, SSW includes MEM$\_$NJIT \citep{2006ApJ...636.1159B,2007SoPh..240..241S}, a Maximum Entropy method; VIS$\_$FWDFIT \citep{2007SoPh..240..241S}, which selects pre-defined source shapes based on their best fitting of visibilities; uv$\_$smooth \citep{2009ApJ...703.2004M}, an interpolation/extrapolation method in the Fourier domain; VIS$\_$CS \citep{2017ApJ...849...10F}, a catalogue-based compressed sensing algorithm; and VIS$\_$WV \citep{2018A&A...615A..59D}, a wavelet-based compressed sensing algorithm. Although each one of these algorithms combines specific values with applicability limitations and specific flaws, a critical comparison of the maps of a given flaring event obtained by the application of all (or most) of these algorithms provides a good picture of what a reliable image of the event could be. 

In particular, MEM$\_$NJIT provides reconstructions characterized by a notable degree of reliability. The capability of fitting the experimental observations is generally satisfactory both in terms of comparison between the predicted and experimental visibility profiles with respect to each {\em{RHESSI}} detector; and in terms of the reduced $\chi^2$ values computed considering either all detectors or just the detectors providing the observations. Further, the algorithm is robust with respect to the level of noise affecting the observations. Moreover, the computational time is among the smallest in the SSW scenario and allows reconstruction within a few seconds. Finally, MEM$\_$NJIT is characterized by super-resolution properties \citep{1985A&A...143...77C}. 

\begin{figure}
\begin{center}
\begin{tabular}{cc}
\includegraphics[height=5.5cm]{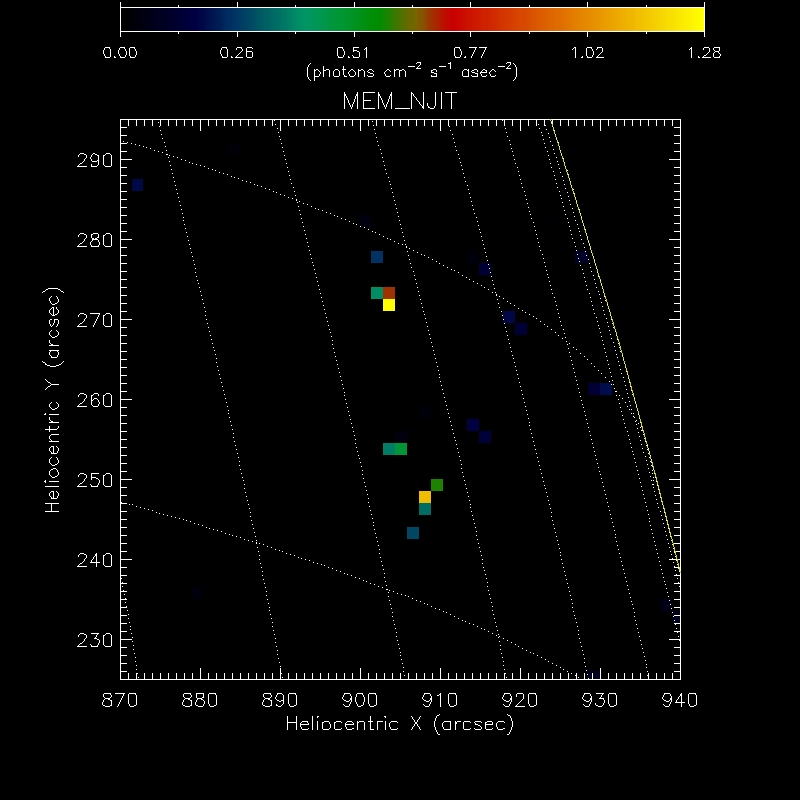} &
\includegraphics[height=5.5cm]{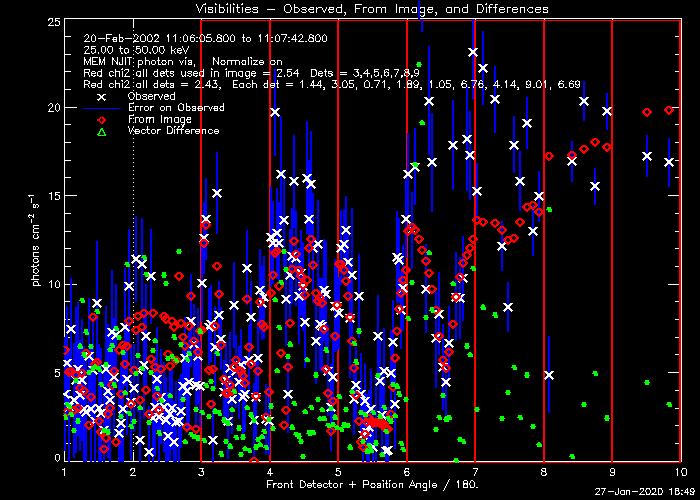}\\
\includegraphics[height=5.5cm]{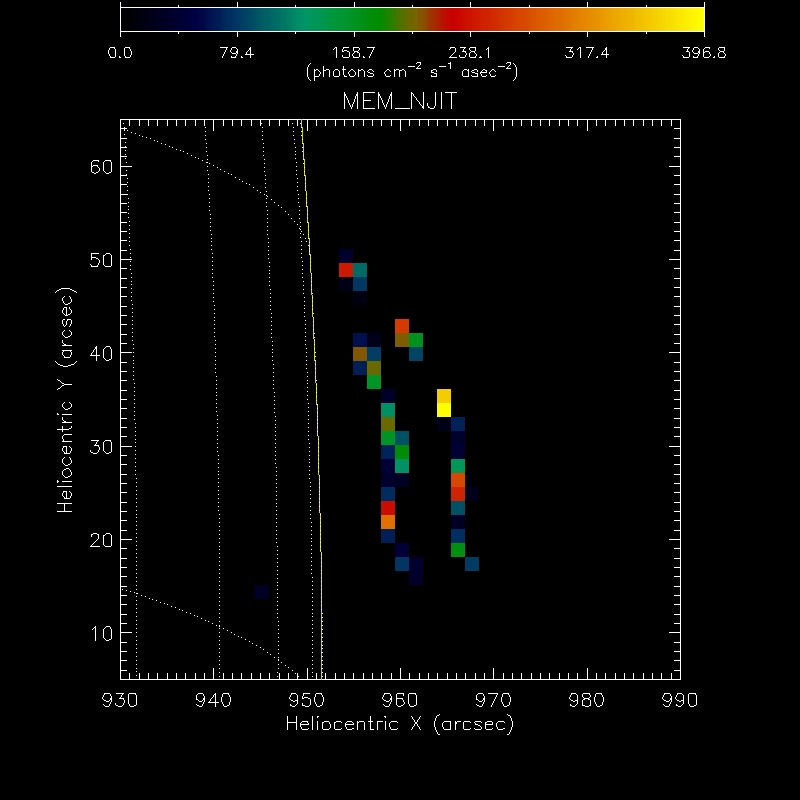} &
\includegraphics[height=5.5cm]{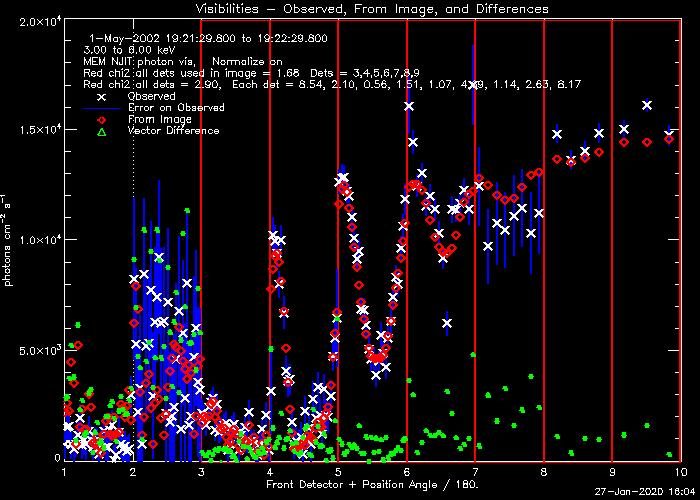} \\
\includegraphics[height=5.5cm]{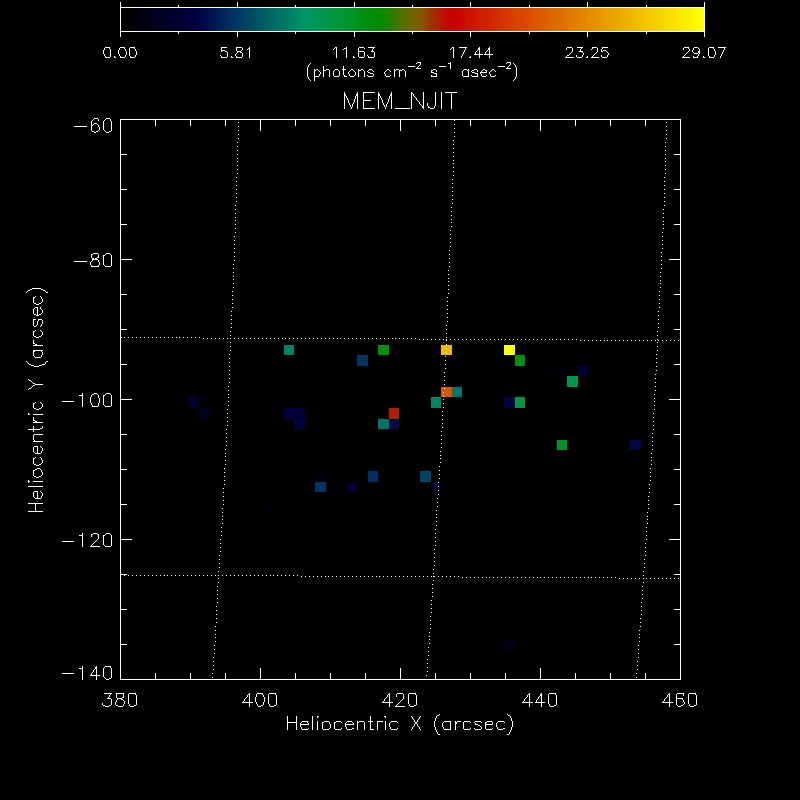} &
\includegraphics[height=5.5cm]{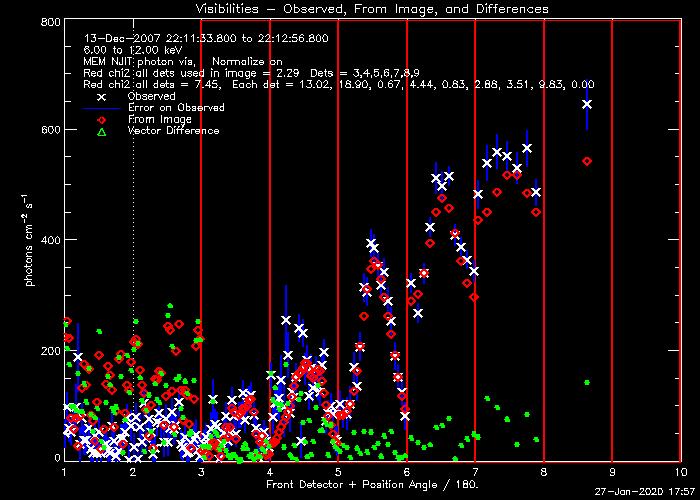}
\end{tabular}
\caption{Three examples of MEM$\_$NJIT misbehaviour. First row: the February 20 2002 event; time interval: 11:06:05 - 11:07:42; energy range: $25-50$ keV; detectors: $3$ through $9$. Second row: the May 1 2002 event; time interval: 19:21:29 - 19:22:29; energy range: $3-6$ keV; detectors: $3$ through $9$. Third row: the December 13 2007 event; time interval: 22:11:33 - 22:12:56; energy range: $6-12$ keV; detectors: $3$ through $9$. Left column: MEM$\_$NJIT reconstructions. Right column: comparison between predicted and measured visibilities}
\label{figure:fig1}
\end{center}
\end{figure}

However, MEM$\_$NJIT sometimes produces images with multiple unrealistically small sources. The origin of this problem is not totally clear but is related to the minimization technique and the setting of the convergence criteria. MEM$\_$NJIT addresses a constrained maximization of the entropy function, which turns into an optimization problem with two penalty terms, the chi-squared function relating the measured and predicted visibilities and a term that ensures the conservation of the overall flux. The optimization of these terms is computationally difficult especially since the optimization functional is not convex, which implies that the numerical schemes may suffer convergence issues. Therefore, the optimization procedure may continue on past the physical meaningful solution to find a solution involving multiple bright points (see Figure \ref{figure:fig1}, where the reconstructed images have dimension $101 \times 101$, the pixel size is $1.5$ arcsec and a zoom is applied for sake of clarity). Of course, this outcome can generally be avoided by increasing the tolerance parameter from its default value (which is currently set at $0.03$) to data-dependent higher values, but this typically results in a worse fitting of the observations and a less accurate estimate of the emission flux and, in general, impedes the use of MEM$\_$NJIT in an automatic pipeline for data processing.  

The present paper illustrates a new algorithm for the constrained maximization of the image entropy which, differently than MEM$\_$NJIT, relies on the optimization of a convex functional. Indeed, this approach aims at the minimization of $\chi^2$ under the constraint of maximum entropy, which leads to the formulation of a convex functional characterized by a single penalty term and, therefore a single regularization parameter that is fixed {\em{a priori}}. The minimization of this functional is performed iteratively and in an alternate fashion \citep{combettes:hal-00643807}: for each iteration, a gradient step minimizes the $\chi^2$ functional and then a proximal step minimizes the negative entropy and projects the result of the first step onto the hyper-surface of vectors with positive components and constant flux. 

The method is implemented in Solar SoftWare (SSW) and can be reached via the HESSI GUI with the name MEM$\_$GE. We point out that the IDL code is implemented in a way that is relatively independent of the instrument providing the experimental visibilities and the related uncertainties. In particular, when using {\em{RHESSI}} visibilities, MEM$\_$GE images are very similar to MEM$\_$NJIT images  when this latter approach works but MEM$\_$GE also provides meaningful images for those cases where the MEM$\_$NJIT images, made with the standard value of the tolerance parameter, are unphysical.

The plan of the paper is as follows. Section 2 provides some details about the formulation of MEM$\_$GE and the optimization algorithm. Section 3 illustrates the results of its application against both experimental visibilities recorded by {\em{RHESSI}} and synthetic visibilities generated within the simulation software of the {\em{Spectrometer Telescope for Imaging X-rays (STIX)}} on-board {\em{Solar Orbiter}}. Comments on these results are contained in Section 4. Our conclusions will be offered in Section 5.

\section{The optimization problem}
Both {\em{RHESSI}} and {\em{STIX}} provide observations, named visibilities, which are calibrated Fourier transforms of the incoming photon flux picked up at specific sampling points ${\mathbf{\xi}}_1,\ldots,{\mathbf{\xi}}_{N_v}$ in the spatial  frequency plane $(u,v)$ (for {\em{RHESSI}}, the number $N_v$ of recorded visibilities is variable and depends on the observation; for {\em{STIX}}, it is fixed at $30$). Therefore image reconstruction for {\em{RHESSI}} and {\em{STIX}} involves the solution of the inverse Fourier transform problem with limited data
\begin{equation}\label{fourier-problem}
    {\mathbf{v}} = {\mathbf{F}}{\mathbf{x}}~, 
\end{equation}
where ${\mathbf{v}} \in \mathbb{C}^{N_v} $ is the vector containing the observed visibilities; the photon flux $N \times N$ image to reconstruct is transformed into the $M$-dimension vector ${\mathbf{x}} \in {\mathbb{R}}^{M}$ with $M= N^2$, by means of the standard column-wise concatenation procedure; ${\mathbf{F}} \in \mathbb{C}^{N_v \times M}$ is the matrix computing the Discrete Fourier Transform (DFT) of ${\mathbf{x}}$ at the spatial frequencies ${\mathbf{\xi}}_1,\ldots,{\mathbf{\xi}}_{N_v}$ sampled by either {\em{RHESSI}} or {\em{STIX}}.

The mathematical basis of MEM$\_$NJIT is the constrained maximization of the entropy
\begin{equation}\label{entropy}
H = - \sum_{j=1}^M x_j \log \frac{x_j}{me}~,
\end{equation}
where $x_j$ is the signal content of pixel $j$, $m=F^{\prime}/M$, $F^{\prime}$ is the total flux in the image and $e$ is the Euler's number ($e=2.71828$). Three constraints are considered in the algorithm; one is
\begin{equation}\label{constraint-1}
    \chi^2 = 0~,
\end{equation}
with
\begin{equation}\label{chi2}
    \chi^2 := \sum_{i=1}^{N_v}\frac{|({\mathbf{F}}{\mathbf{x}})_i - v_i|^2}{\sigma^2_i} - N_v~,
\end{equation}
where ${\mathbf{\sigma}}$ is the vector of the experimental uncertainties. A second constraint involves the flux and is given by
\begin{equation}\label{constraint-2}
    F = 0~,
\end{equation}
with 
\begin{equation}\label{flux}
F = \sum_{j=1}^M x_j -F^{\prime}~.
\end{equation}
Finally, a third constraint requires that all components of ${\mathbf{x}}$ must be non-negative. The algorithm implemented in the MEM$\_$NJIT IDL code addresses the constrained maximum problem
\begin{equation}\label{max_MEM-NJIT}
\arg\max_{{\mathbf{x}} \geq 0} \{H - \alpha \chi^2 - \beta F\}~,
\end{equation}
where $\alpha$ and $\beta$ are the Lagrange multipliers associated to constraints (\ref{constraint-1}) and (\ref{constraint-2}); these two parameters are not estimated {\em{a priori}} but are updated during the maximization process. The first main drawback of this approach is that the maximization problem (\ref{max_MEM-NJIT}) involves a functional which is not convex and therefore numerical schemes may lead to unstable solutions. Further, the functional is characterized by two Lagrange multipliers, whose updating process is sometimes non optimal. When one of these conditions occurs, MEM$\_$NJIT produces unphysical reconstructions like the ones shown in Figure \ref{figure:fig1}.

MEM$\_$GE addresses these two issues by providing a different formulation of the maximum entropy optimization problem. The first idea is to replace the maximization problem ({\ref{max_MEM-NJIT}}) with the minimization problem
\begin{equation}\label{min_MEM-GE}
\arg\min_{{\mathbf{x}} \geq 0} \{\chi^2 - \lambda H\}~,
\end{equation}
under the flux constraint (\ref{constraint-2}) and where $\lambda$ is the regularization parameter. The main advantage of this approach is that now the optimization problem is convex and therefore it can be addressed by relying on several numerical methods whose convergence properties are well-established. In particular, in MEM$\_$GE we adopted the following iterative scheme whereby, at each iteration we compute \citep{combettes:hal-00643807}
\begin{enumerate}
\item A gradient step to minimize $\chi^2$;
\item A proximal step to maximize the entropy subjected to the positivity and flux constraints.
\end{enumerate}

After this second step, the algorithm is accelerated by computing a linear combination with the approximation corresponding to the previous iteration and a monotonicity check is also performed \citep{beck2009fast}. 

Two main technical aspects are involved by the implementation of the algorithm. First, the regularization parameter $\lambda$ is {\em{a priori}} determined relying on the observation that an over-regularizing value $\lambda_0$ of this parameter implies that the corresponding regularized solution must be related to the average flux by
\begin{equation}\label{lambda-1}
\left| \frac{({\mathbf{x}}_{\lambda_0})_j - m}{m}\right| \leq 1
\end{equation}
for all $j$. Simple numerical approximations show that (\ref{lambda-1}) implies
\begin{equation}\label{lambda-2}
\lambda_0 \geq  \max_j | \partial_j \chi^2 ({\mathbf{x}}_{\lambda_0})| %|(F_{\sigma}^T(F_{\sigma}{\mathbf{x}}_{\lambda_0} - {\mathbf{v}}_{\sigma}))_j|~,
\end{equation}
where $\partial_j$ indicates the partial derivative along the $j$-th direction. In order to determine $\lambda_0$ we approximate each component of ${\mathbf{x}}_{\lambda_0}$ with $m$. This results in a over-estimate of the regularization parameter; therefore the optimal value for the regularization parameter is chosen as a rate of the estimated $\lambda_0$, where this rate is determined as a function of the visibility signal-to-noise ratio using a heuristic look-up table.
%\begin{equation}\label{lambda-3}
%(F_{\sigma})_{ij}=\frac{F_{ij}}{\sigma_i}~~~~~({\mathbf{v}}_{\sigma})_i = \frac{{\mathbf{v}}_i}{\sigma_i}~.
%\end{equation}

Second, the realization of the flux constraint in the second step relies on the solution of the not-regularized constrained minimum problem
\begin{equation}\label{flux-constraint}
\arg\min_{{\mathbf{x}} \geq 0} \chi^2~,
\end{equation}
which is performed by applying the projected Landweber method \citep{piana1997projected}: the estimate of the flux is computed by summing up the pixel content of the solution of problem (\ref{flux-constraint}).

%  addresses the problem of minimizing the entropy under the simultaneous assumptions that the reconstructed image is made of pixels with positive contents, which perfectly predict the experimental data in such a way that the overall flux is conserved. In computational terms, this requires the numerical solution of a maximization problem characterized by two regularization parameters, corresponding to two Lagrange multipliers, i.e. $\chi^2$ and the photon flux. The iterative scheme realizing the optimization requires, at each step, the update of both the parameters and when  

%\section{Optimization issues}

\section{Application to X-ray visibilities}
This section validates the reliability of MEM$\_$GE in the case of both observations provided by {\em{RHESSI}} and synthetic visibilies simulated according to the {\em{STIX}} imaging concept setup. 

\subsection{RHESSI}
\begin{figure}
\begin{center}
\begin{tabular}{cc}
\includegraphics[height=5.5cm]{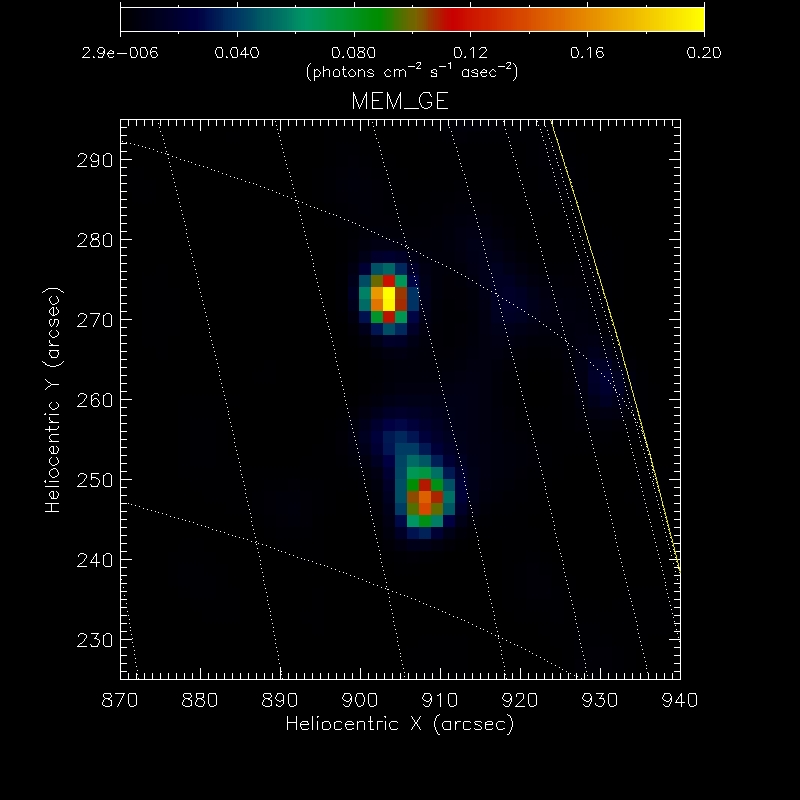} &
\includegraphics[height=5.5cm]{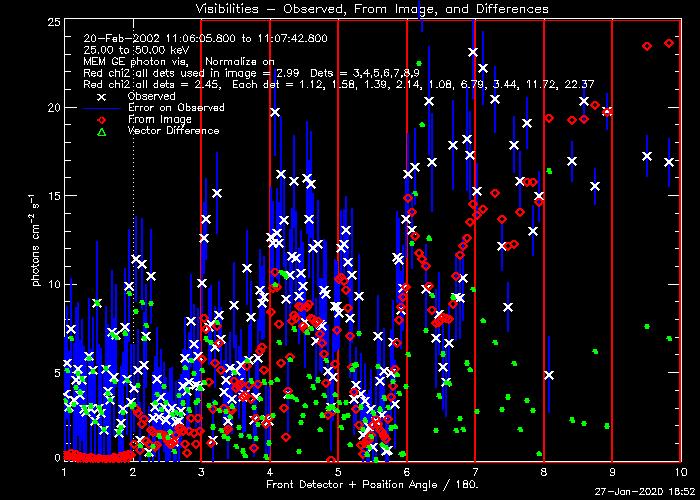}\\
\includegraphics[height=5.5cm]{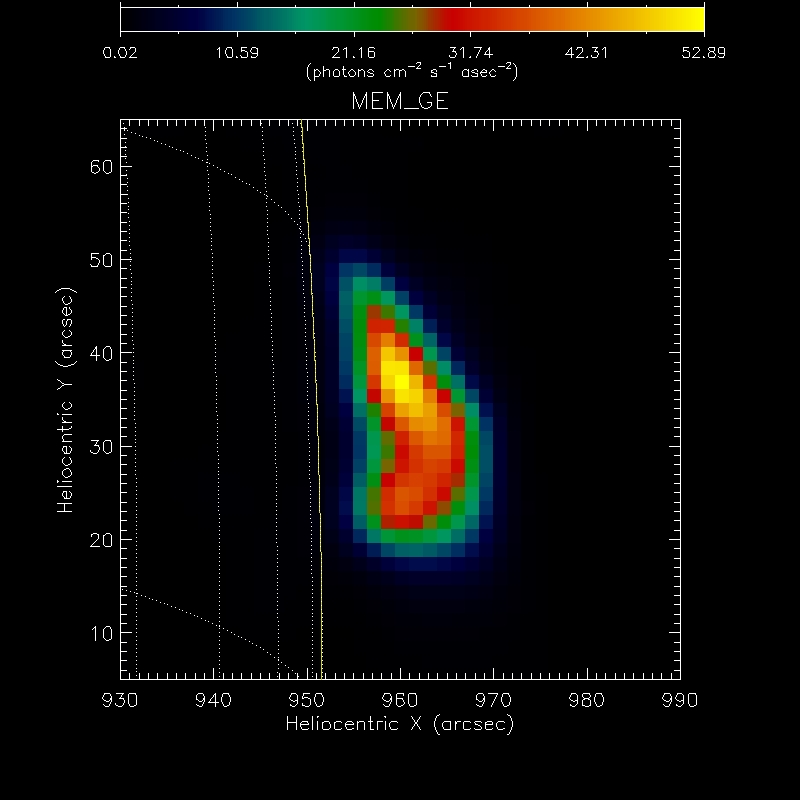} &
\includegraphics[height=5.5cm]{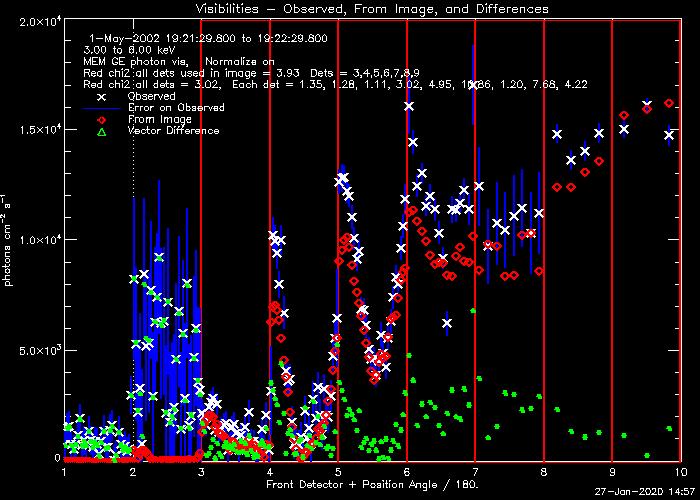} \\
\includegraphics[height=5.5cm]{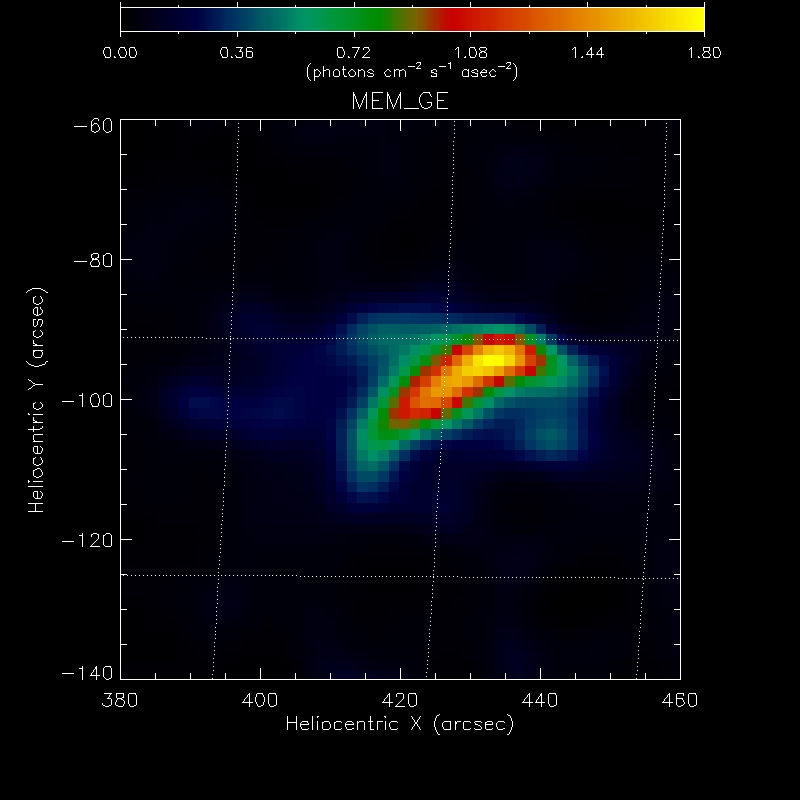} &
\includegraphics[height=5.5cm]{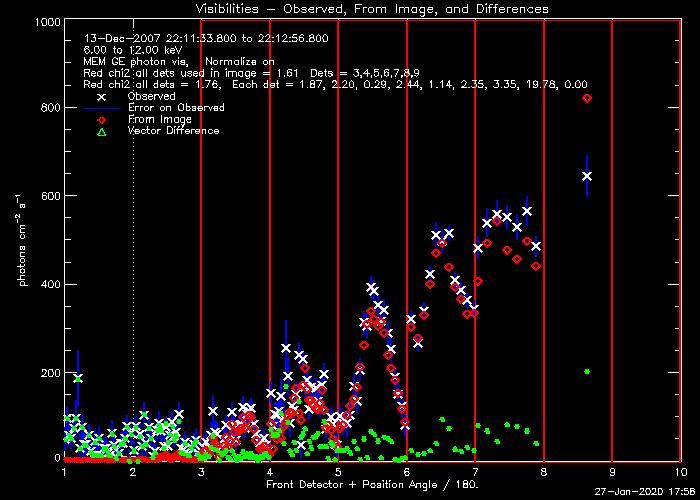}
\end{tabular}
\caption{Left: MEM$\_$GE reconstructions (left) and corresponding comparisons between predicted and observed visibilities (right) for the same cases as in Figure {\ref{figure:fig1}}.}
\label{figure:fig2}
\end{center}
\end{figure}

\begin{figure}[h]
\begin{tabular}{ccc}
    \includegraphics[width=0.3\textwidth]{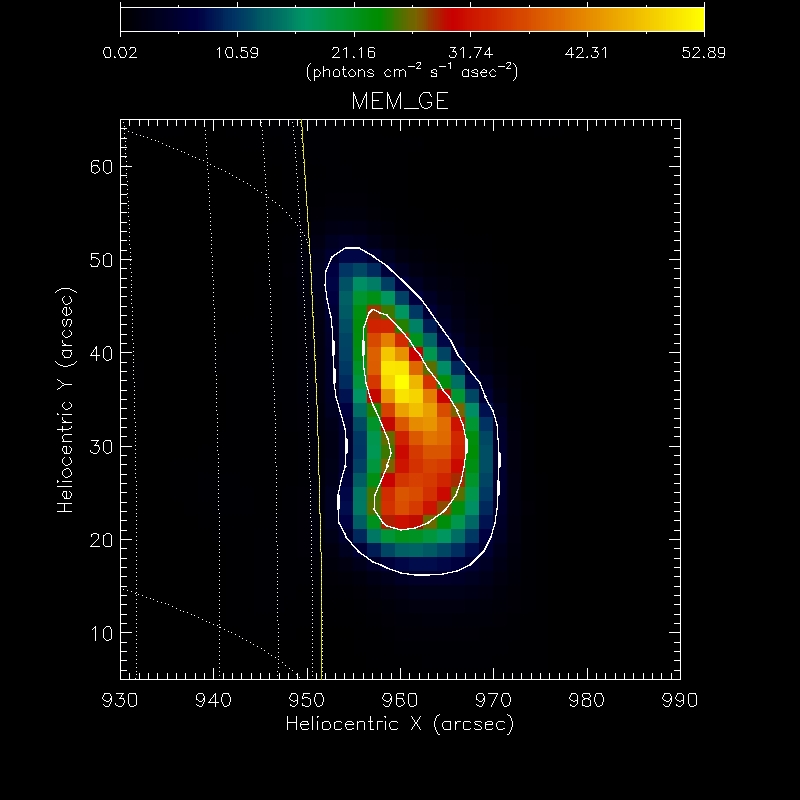}
    &\includegraphics[width=0.3\textwidth]{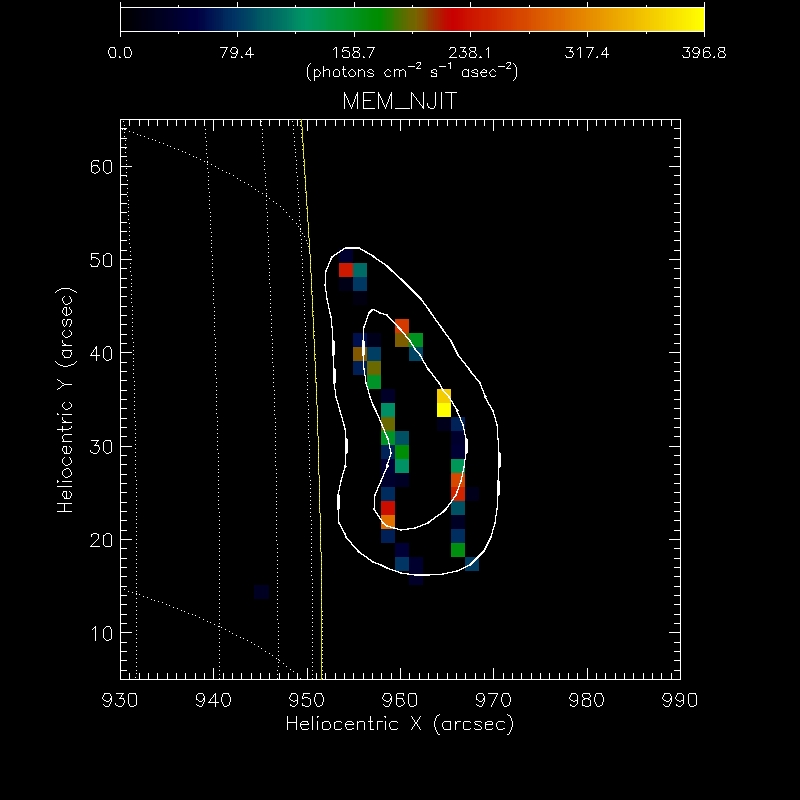}
    &\includegraphics[width=0.3\textwidth]{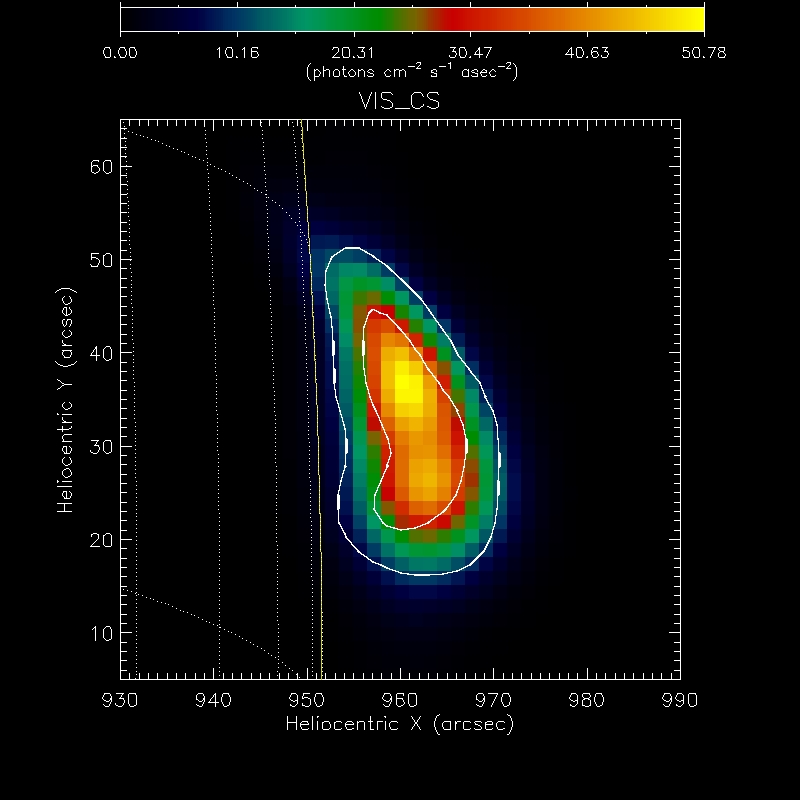}\\
    MEM$\_$GE & MEM$\_$NJIT & VIS$\_$CS \\
    \includegraphics[width=0.3\textwidth]{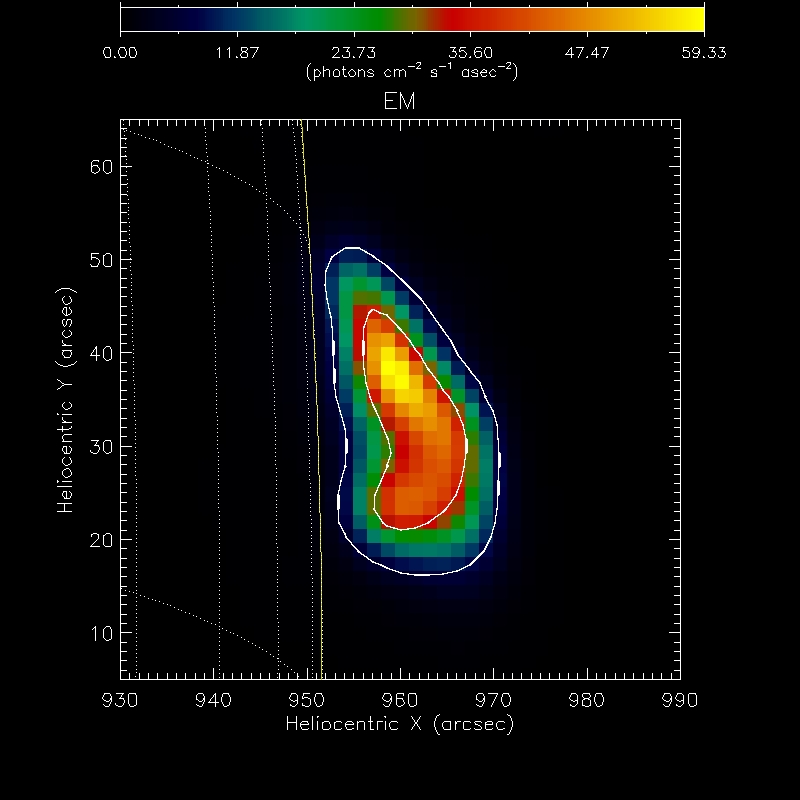}
    &\includegraphics[width=0.3\textwidth]{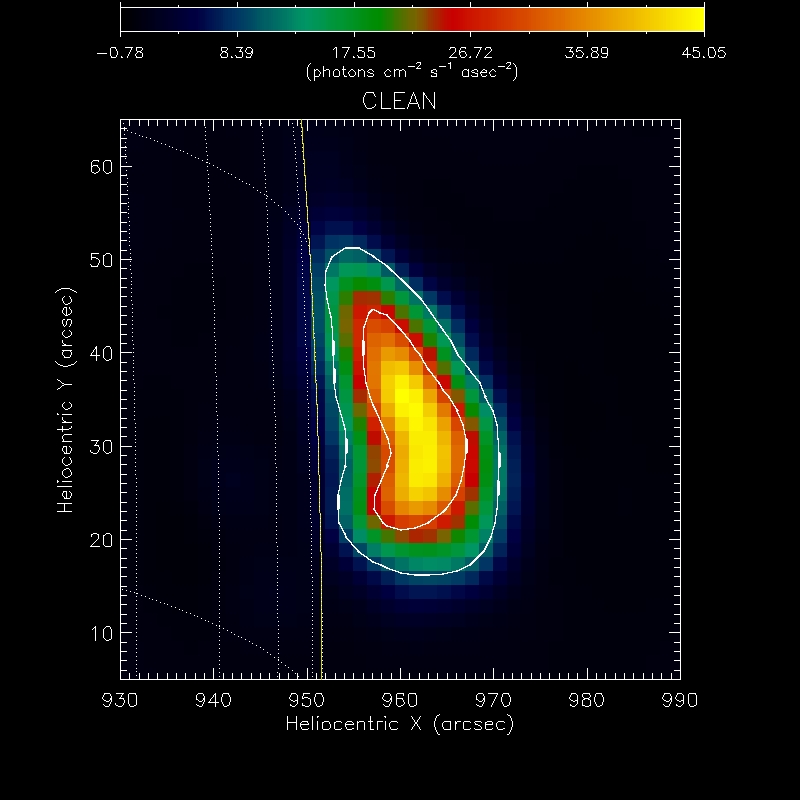} 
    &\includegraphics[width=0.3\textwidth]{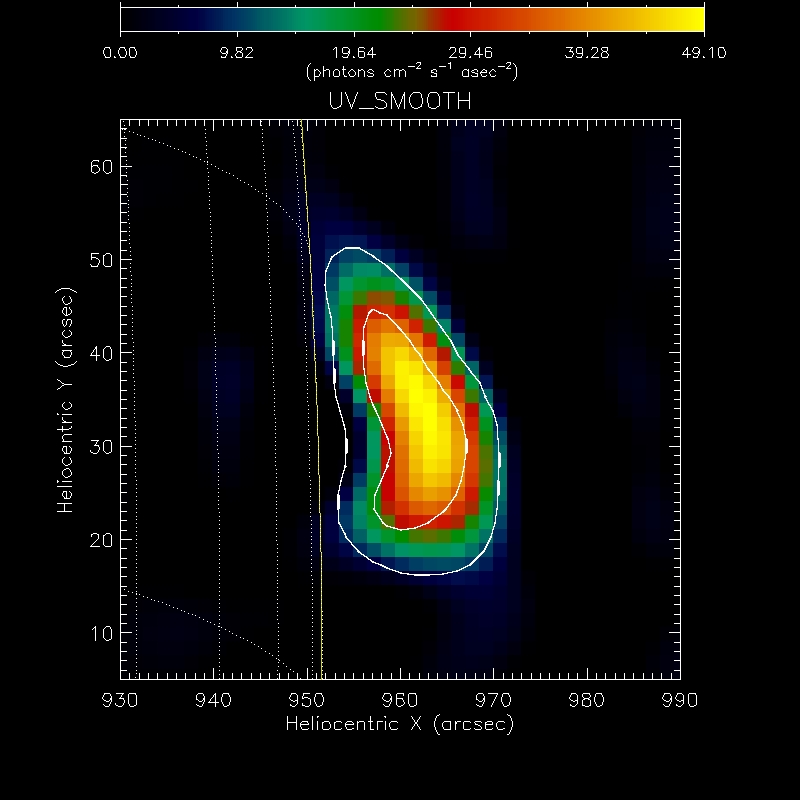} \\
    EM & Clean & uv$\_$smooth
\end{tabular}
    \caption{Reconstructions of the May 1 2002 event provided by six imaging methods available from the {\em{HESSI}} GUI. The observations and conditions are the same as in Figure \ref{figure:fig1} for this same event. For sake of comparison, contour levels of MEM$\_$GE corresponding to $10\%$ and $50\%$ of the maximum intensity are superimposed to the reconstructions.}
    \label{figure:fig3}
\end{figure}

\begin{figure}[h]
\begin{tabular}{ccc}
    \includegraphics[width=0.3\textwidth]{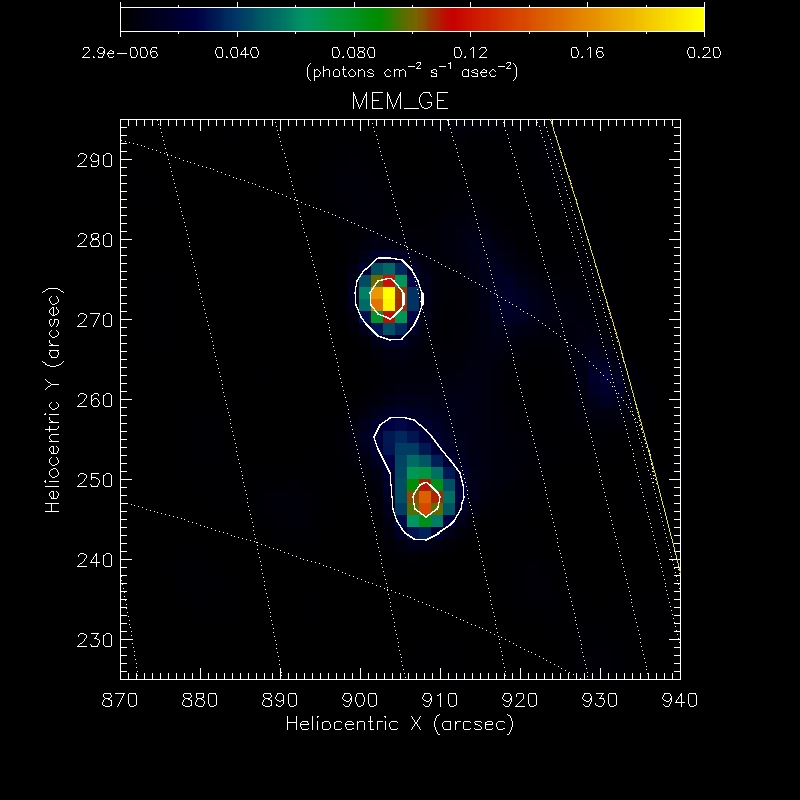}
    &\includegraphics[width=0.3\textwidth]{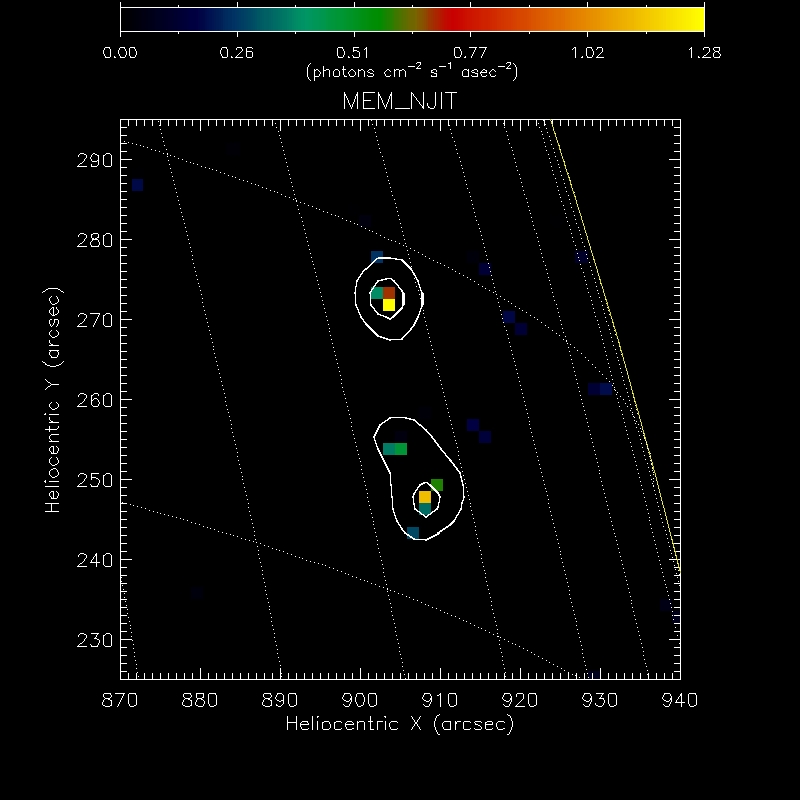}
    &\includegraphics[width=0.3\textwidth]{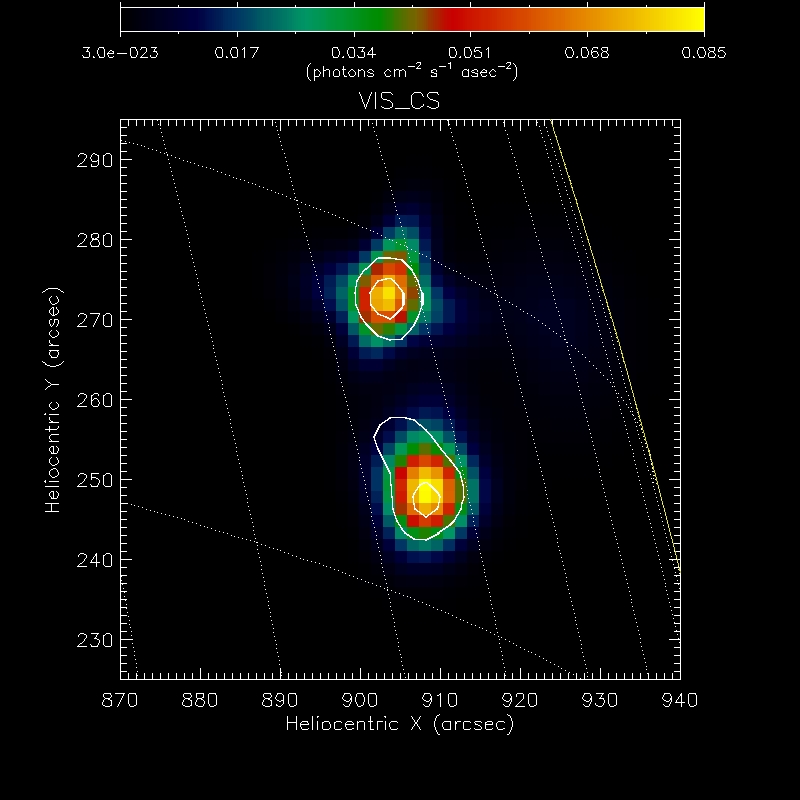}\\
      MEM$\_$GE & MEM$\_$NJIT & VIS$\_$CS \\
    \includegraphics[width=0.3\textwidth]{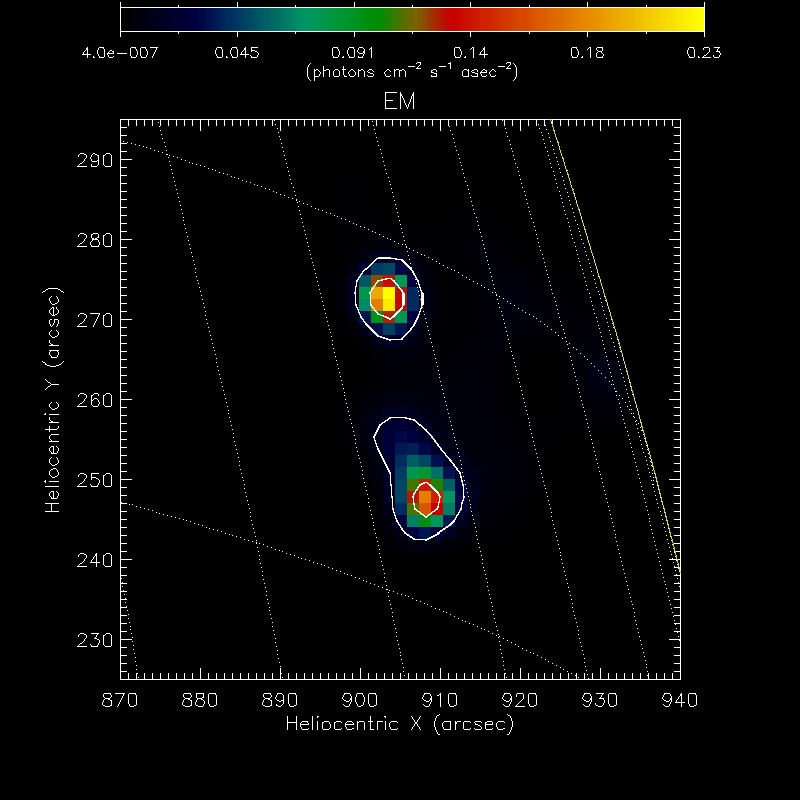}
    &\includegraphics[width=0.3\textwidth]{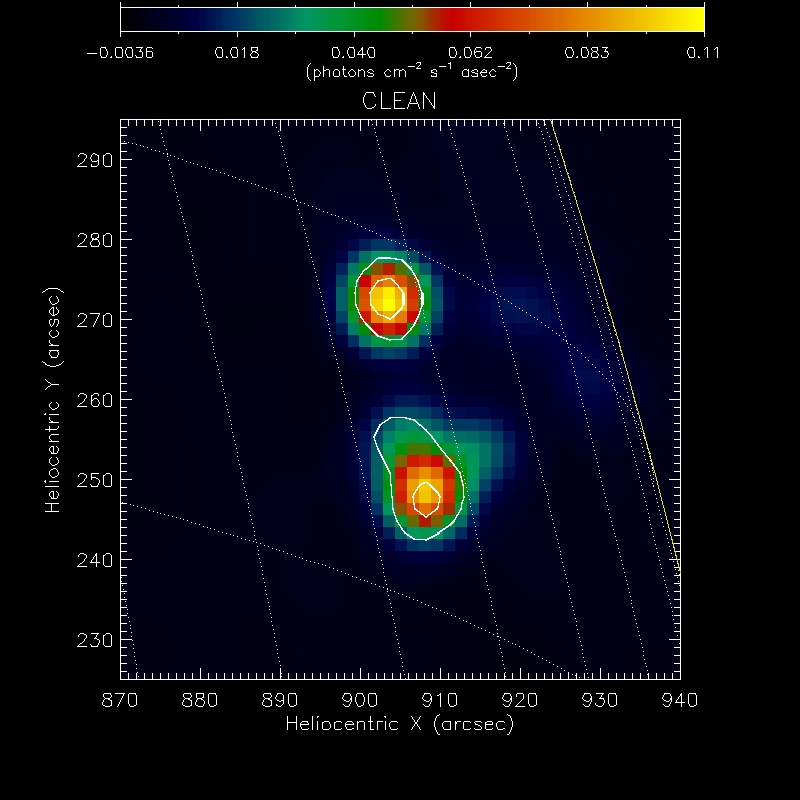} 
    &\includegraphics[width=0.3\textwidth]{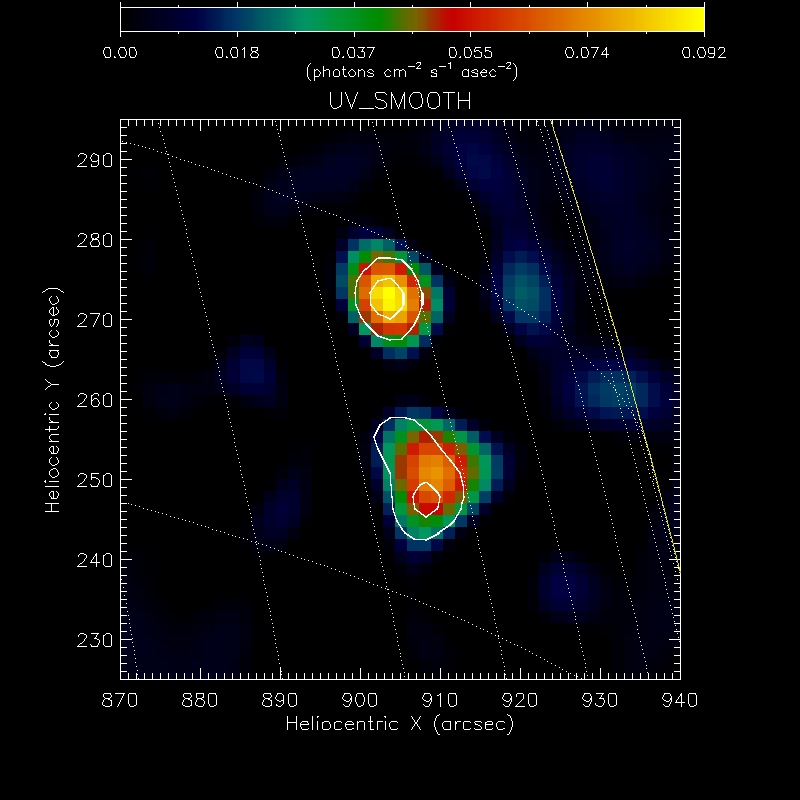}\\
        EM & Clean & uv$\_$smooth
\end{tabular}
    \caption{Reconstructions of the February 20 2002 event provided by six imaging methods available from the {\em{HESSI}} GUI. The observations and conditions are the same as in Figure \ref{figure:fig1} for this same event. For sake of comparison, contour levels of MEM$\_$GE corresponding to $10\%$ and $50\%$ of the maximum intensity are superimposed to the reconstructions.}    \label{figure:fig4}
\end{figure}

\begin{figure}[h]
\begin{tabular}{ccc}
    \includegraphics[width=0.3\textwidth]{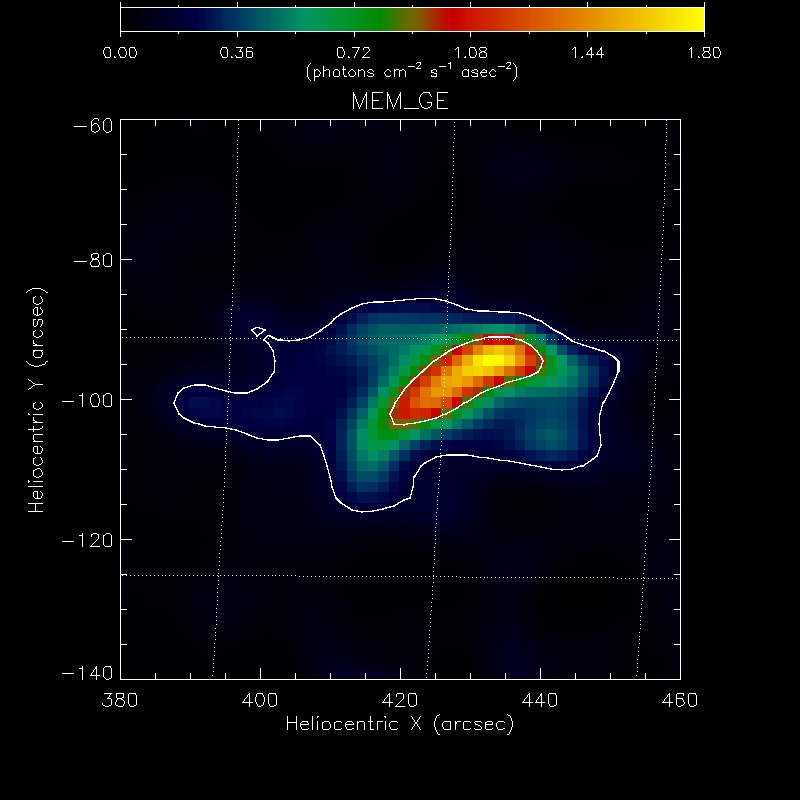}
    &\includegraphics[width=0.3\textwidth]{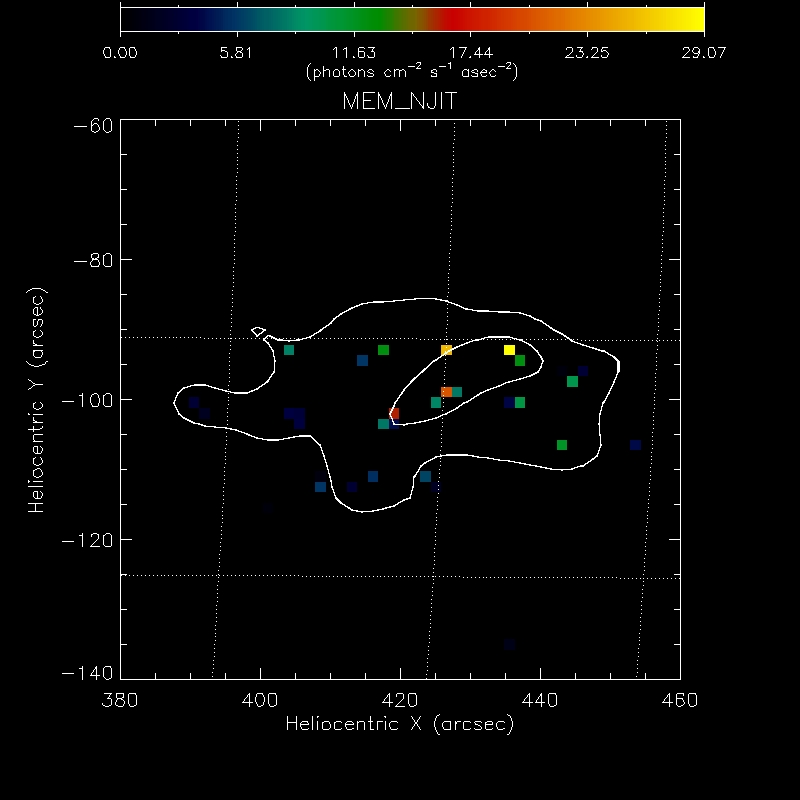}
    &\includegraphics[width=0.3\textwidth]{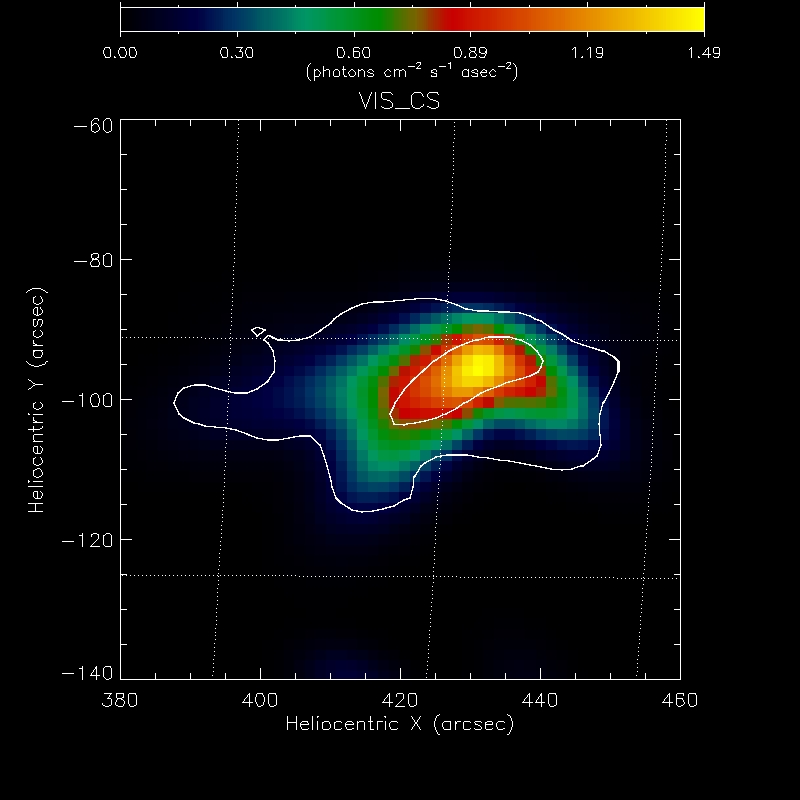}\\
        MEM$\_$GE & MEM$\_$NJIT & VIS$\_$CS \\
    \includegraphics[width=0.3\textwidth]{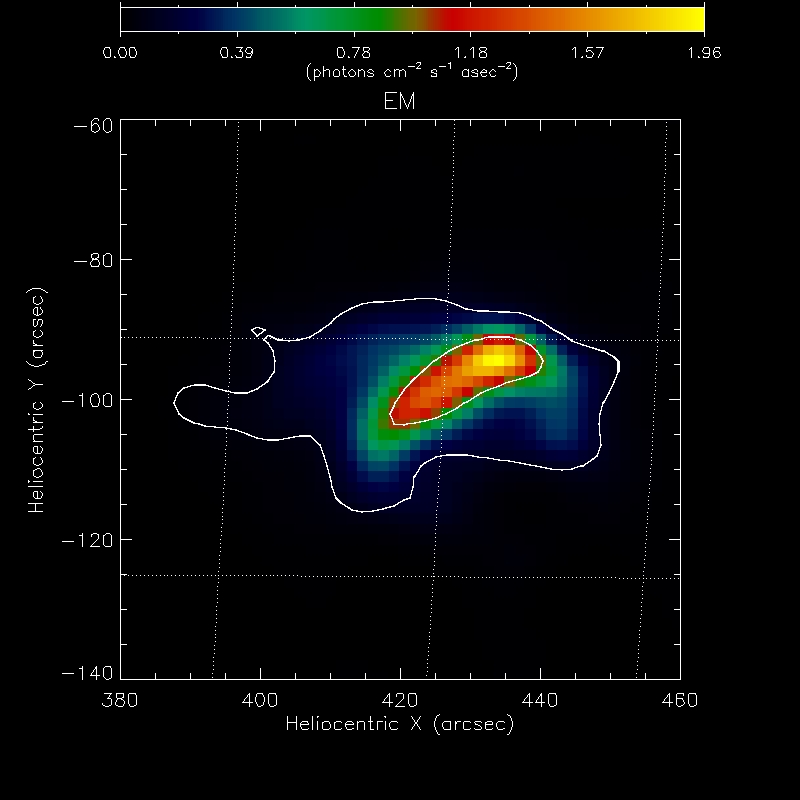}
    &\includegraphics[width=0.3\textwidth]{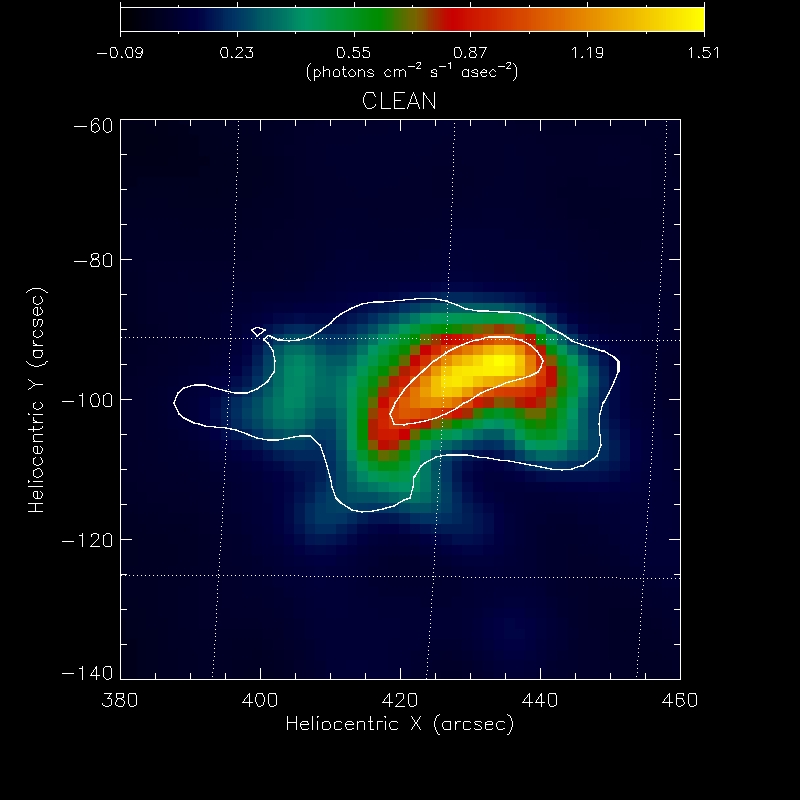} 
    &\includegraphics[width=0.3\textwidth]{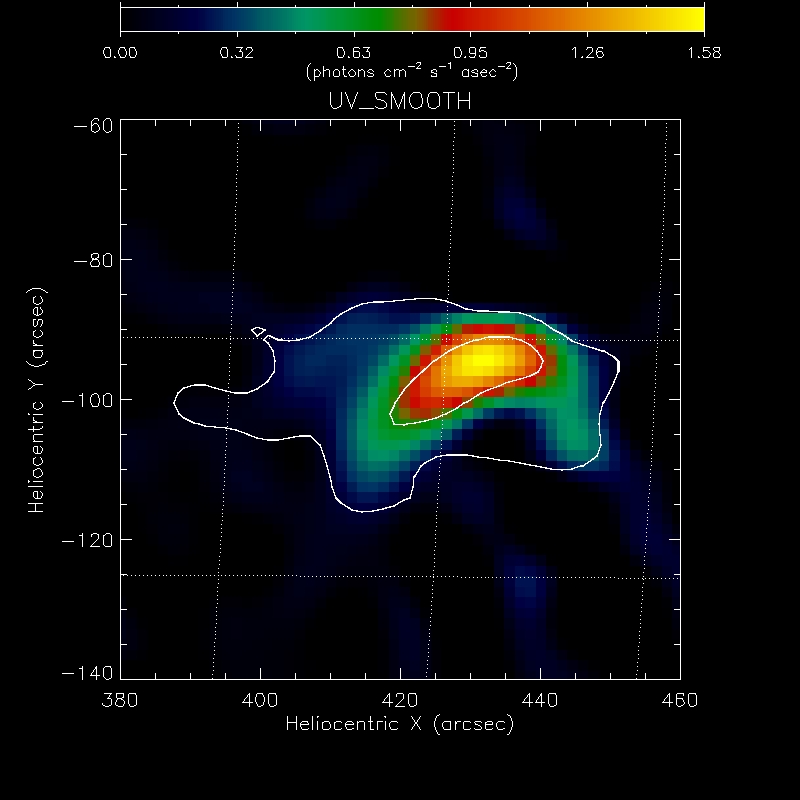}\\
            EM & Clean & uv$\_$smooth
\end{tabular}
    \caption{Reconstructions of the December 13 2007 event provided by six imaging methods available from the {\em{HESSI}} GUI. The observations and conditions are the same as in Figure \ref{figure:fig1} for this same event. For sake of comparison, contour levels of MEM$\_$GE corresponding to $10\%$ and $50\%$ of the maximum intensity are superimposed to the reconstructions.}    \label{figure:fig5}
\end{figure}

\begin{figure}[h]
\begin{tabular}{ccc}
    \includegraphics[width=0.3\textwidth]{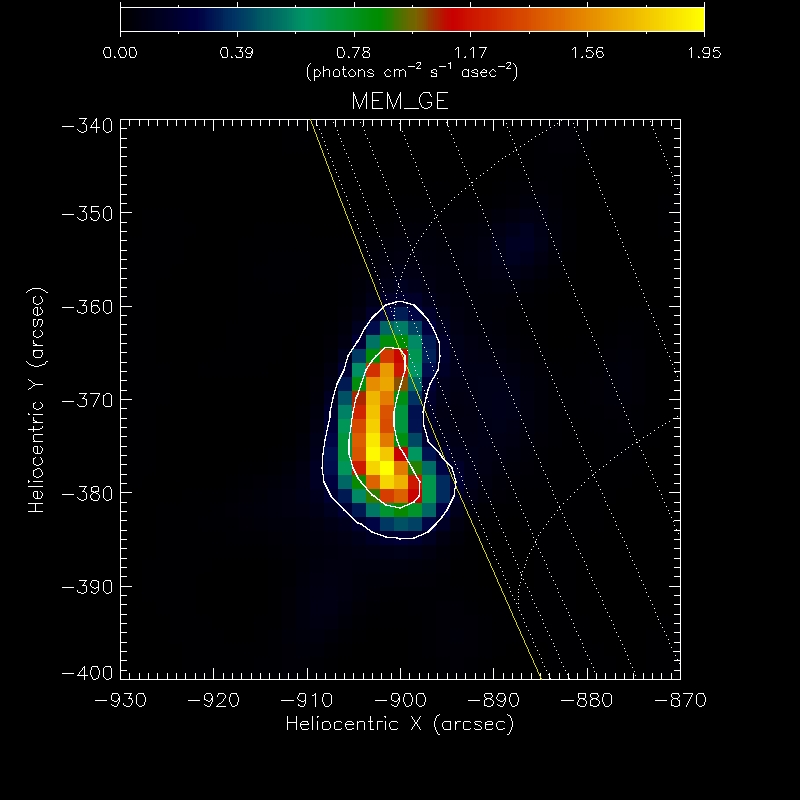}
    &\includegraphics[width=0.3\textwidth]{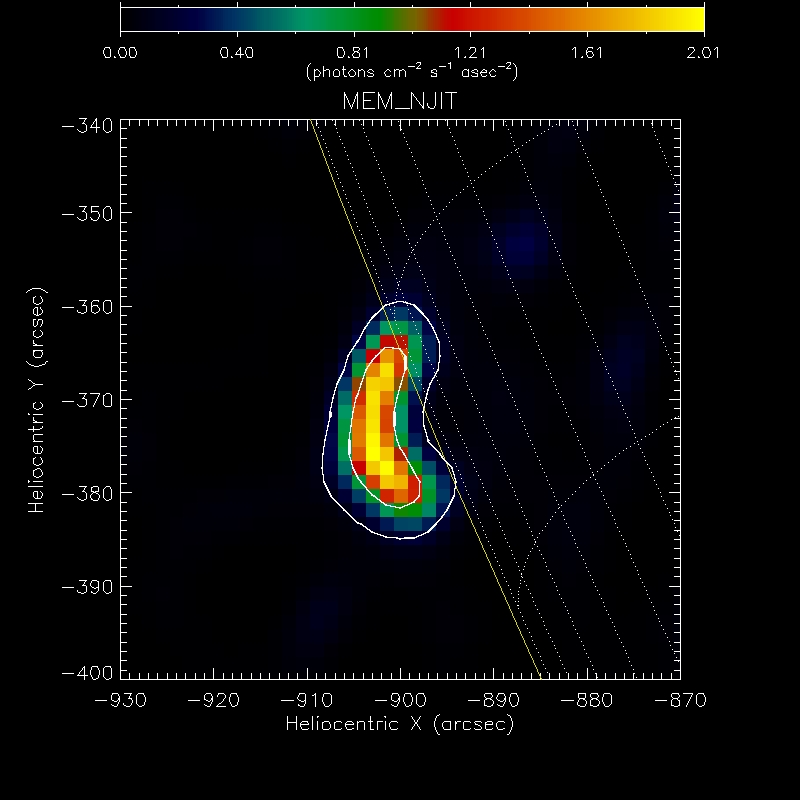}
    &\includegraphics[width=0.3\textwidth]{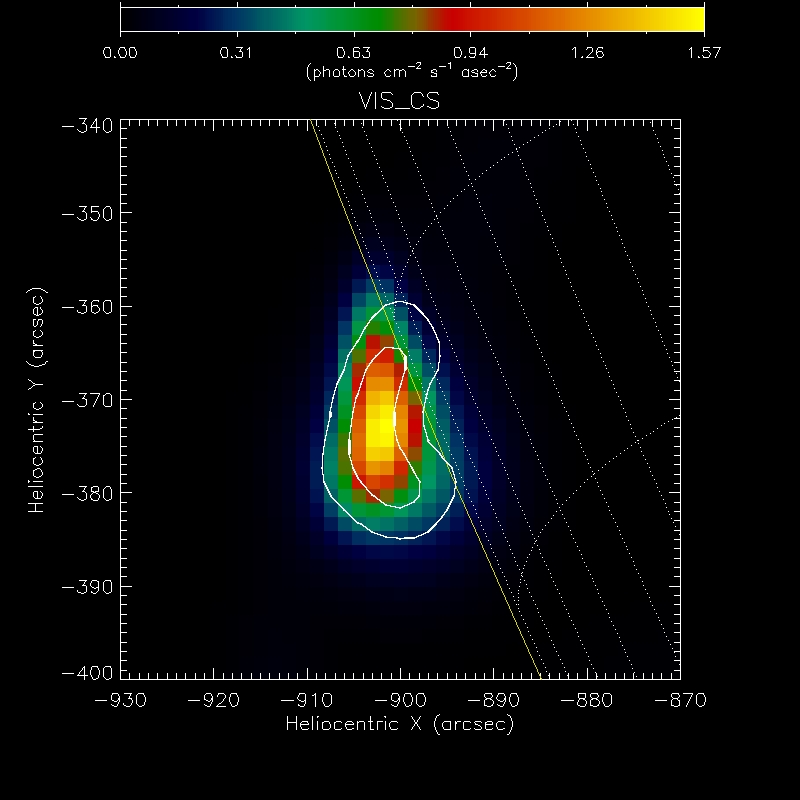}\\
            MEM$\_$GE & MEM$\_$NJIT & VIS$\_$CS \\
    \includegraphics[width=0.3\textwidth]{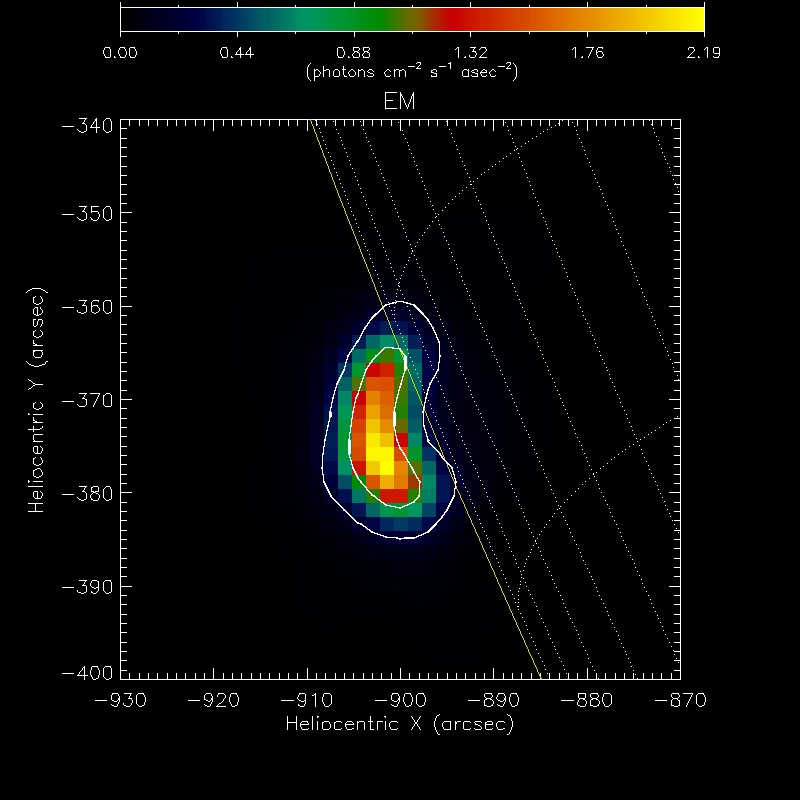}
    &\includegraphics[width=0.3\textwidth]{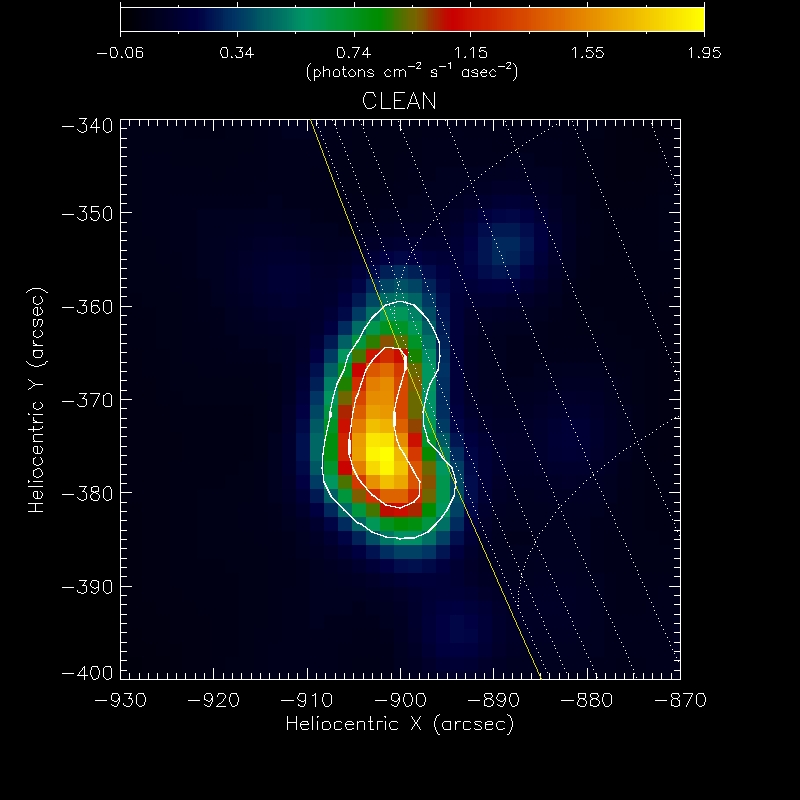} 
    &\includegraphics[width=0.3\textwidth]{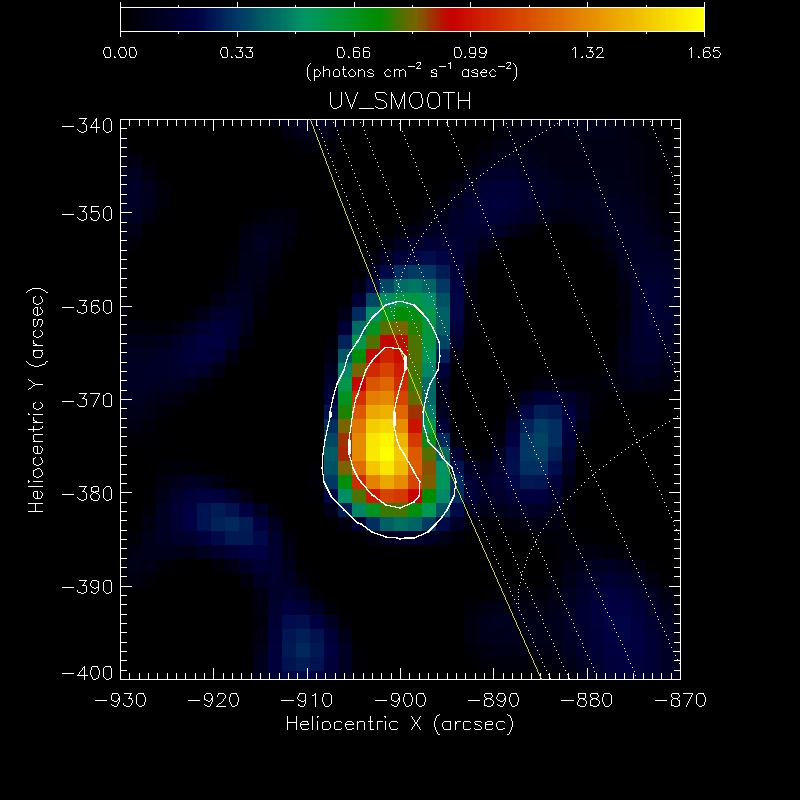}\\
                EM & Clean & uv$\_$smooth
\end{tabular}
    \caption{Reconstructions of the February 13 2002 event provided by the same six imaging methods considered in Figures \ref{figure:fig3} through \ref{figure:fig5}. Time interval: 12:29:40 - 12:31:22; energy range: $6-12$ keV; detectors: $3$ through $9$. For sake of comparison, contour levels of MEM$\_$GE corresponding to $10\%$ and $50\%$ of the maximum intensity are superimposed to the reconstructions.}
    \label{figure:fig6}
\end{figure}

\begin{figure}[h]
\begin{tabular}{ccc}
    \includegraphics[width=0.3\textwidth]{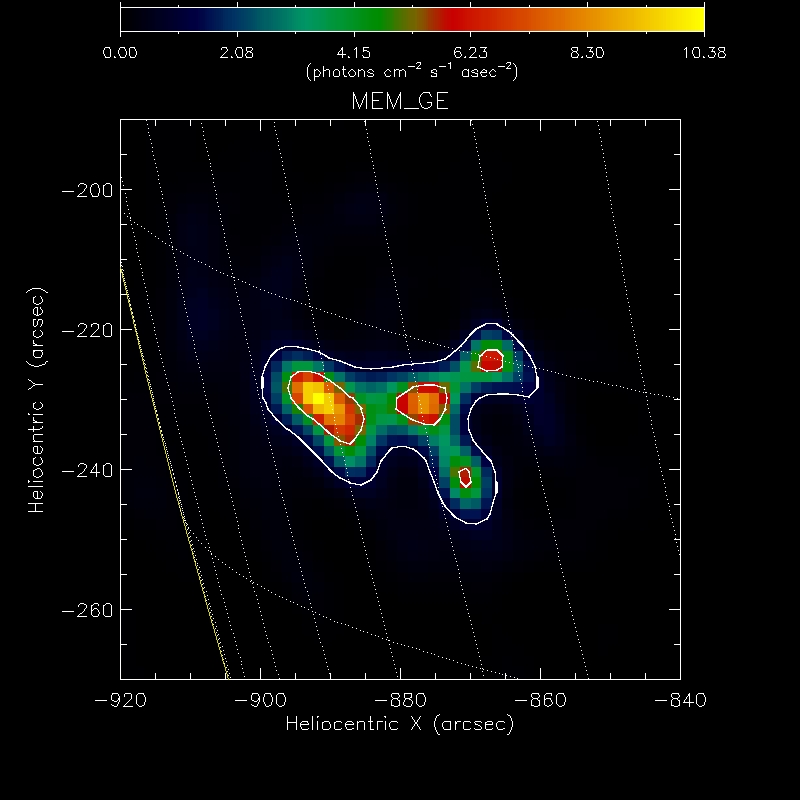}
    &\includegraphics[width=0.3\textwidth]{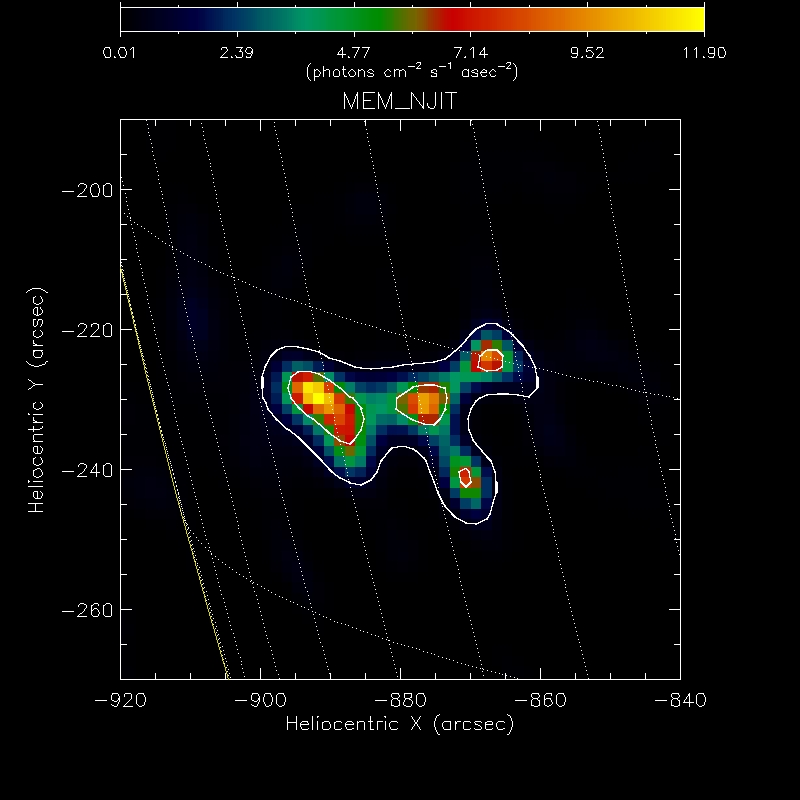}
    &\includegraphics[width=0.3\textwidth]{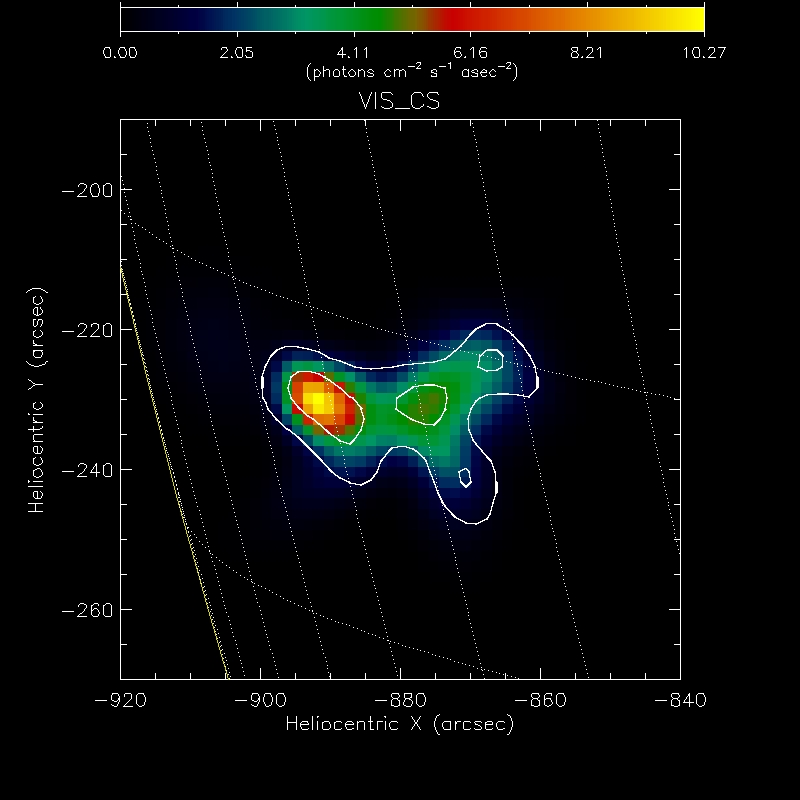}\\
                MEM$\_$GE & MEM$\_$NJIT & VIS$\_$CS \\
    \includegraphics[width=0.3\textwidth]{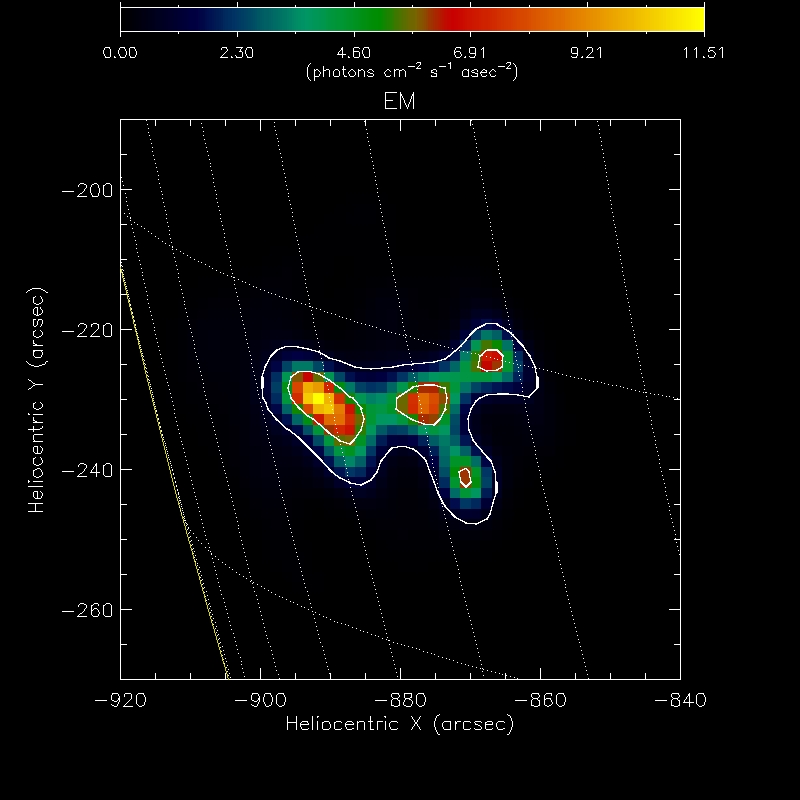}
    &\includegraphics[width=0.3\textwidth]{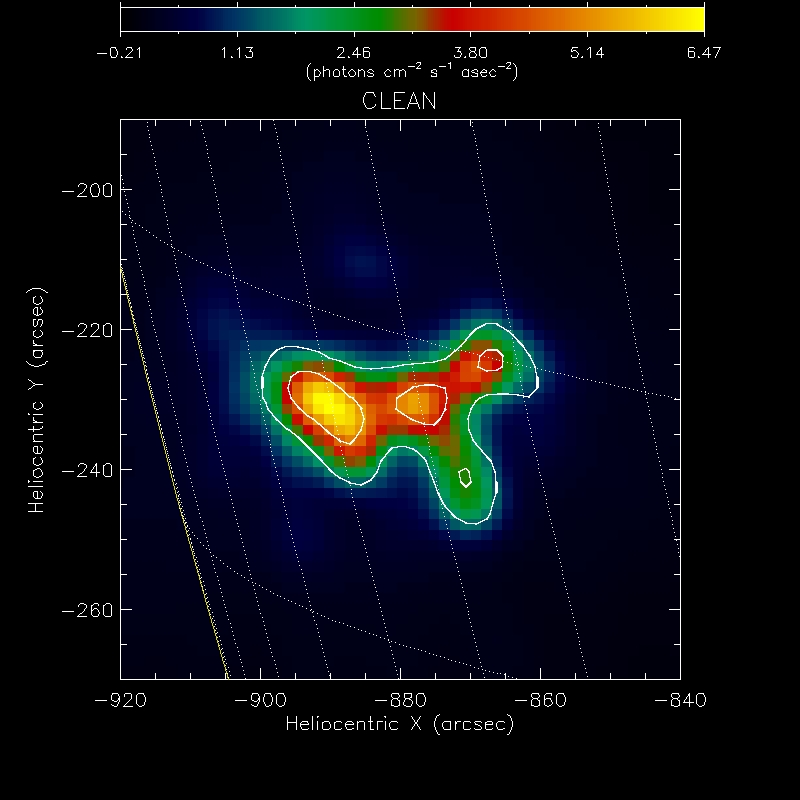} 
    &\includegraphics[width=0.3\textwidth]{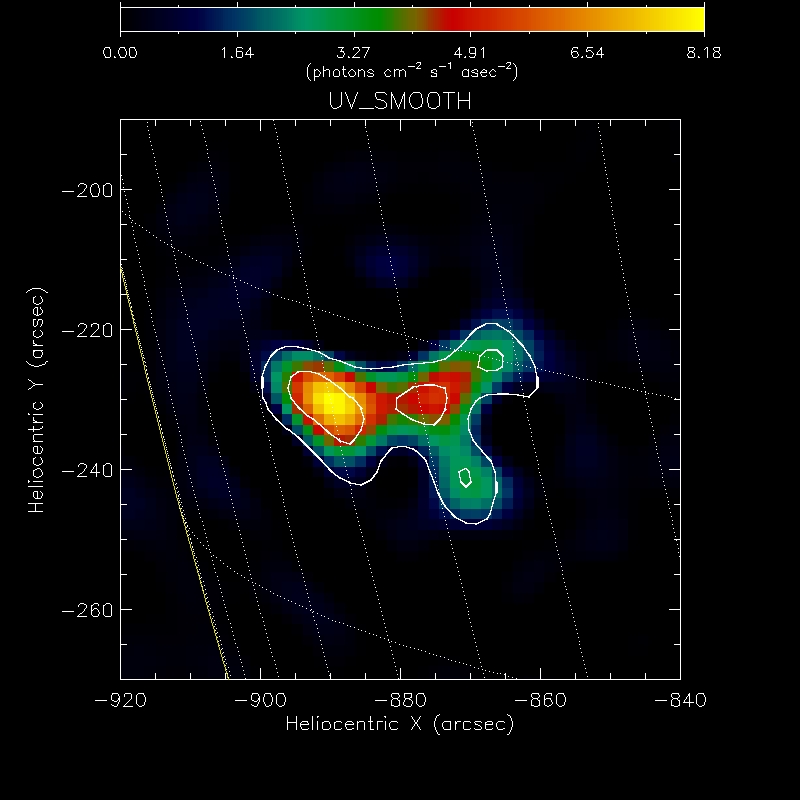}\\
                    EM & Clean & uv$\_$smooth
\end{tabular}
    \caption{Reconstructions of the July 23 2002 event provided by the same six imaging methods considered in Figures \ref{figure:fig3} through \ref{figure:fig5}. Time interval: 00:29:23 - 00:29:39; energy range: $25-50$ keV; detectors: $3$ through $9$. For sake of comparison, contour levels of MEM$\_$GE corresponding to $10\%$ and $50\%$ of the maximum intensity are superimposed to the reconstructions.}
    \label{figure:fig7}
\end{figure}

\begin{figure}[h]
\begin{tabular}{ccc}
    \includegraphics[width=0.3\textwidth]{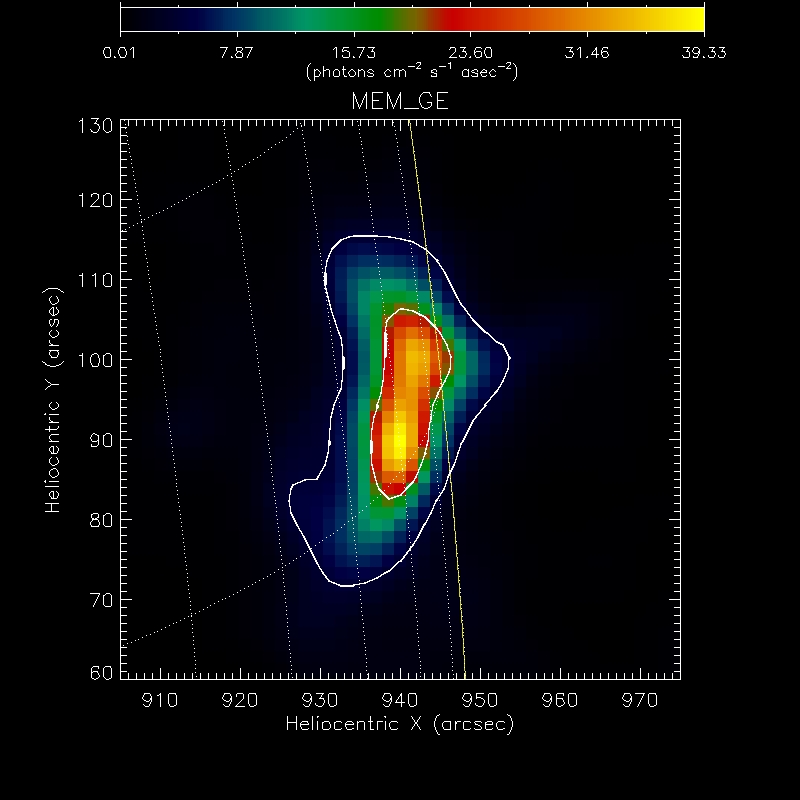}
    &\includegraphics[width=0.3\textwidth]{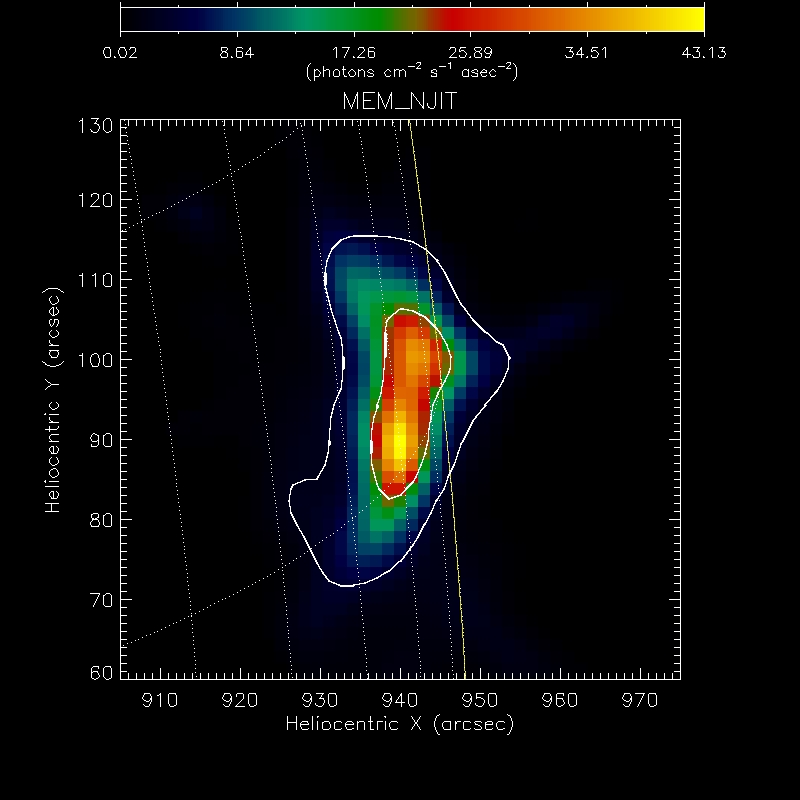}
    &\includegraphics[width=0.3\textwidth]{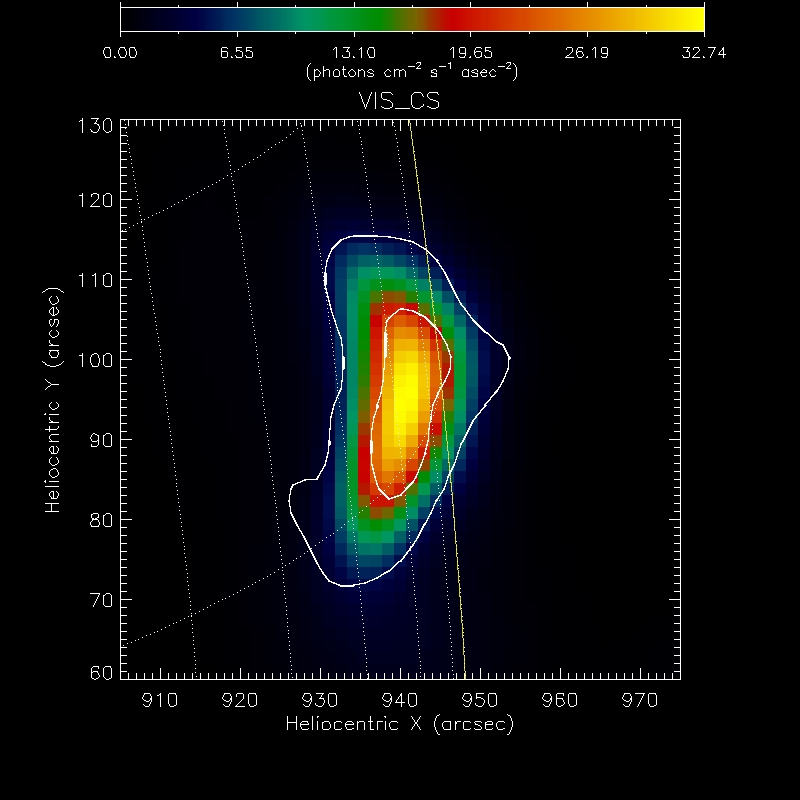}\\
                    MEM$\_$GE & MEM$\_$NJIT & VIS$\_$CS \\
    \includegraphics[width=0.3\textwidth]{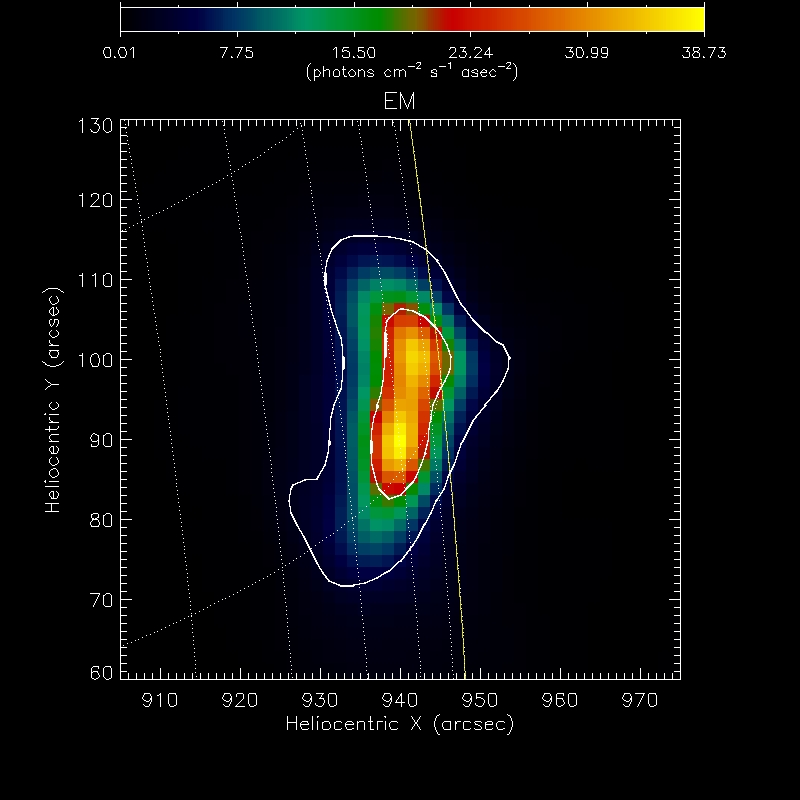}
    &\includegraphics[width=0.3\textwidth]{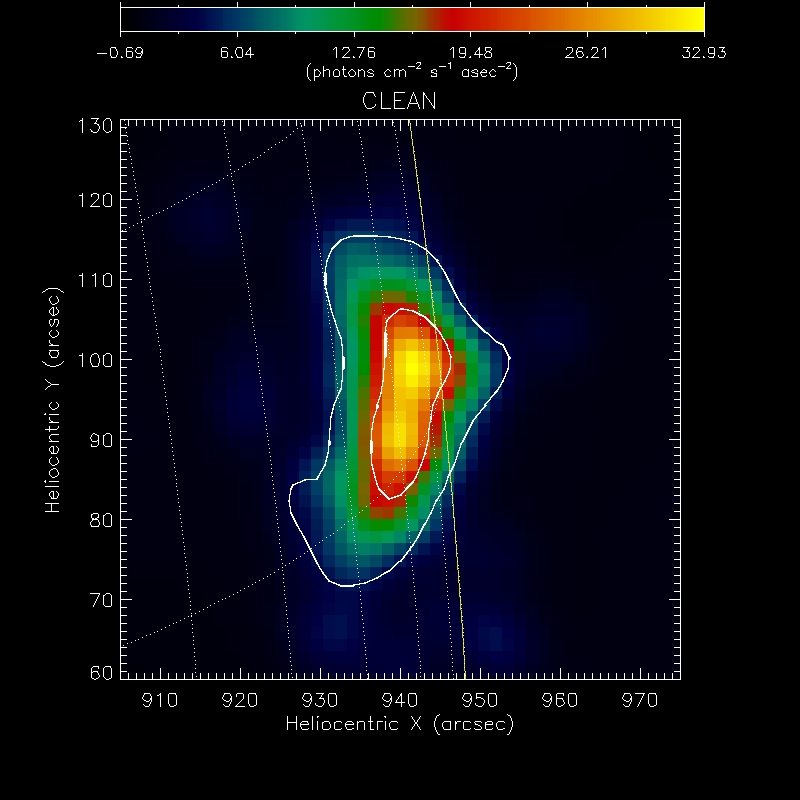} 
    &\includegraphics[width=0.3\textwidth]{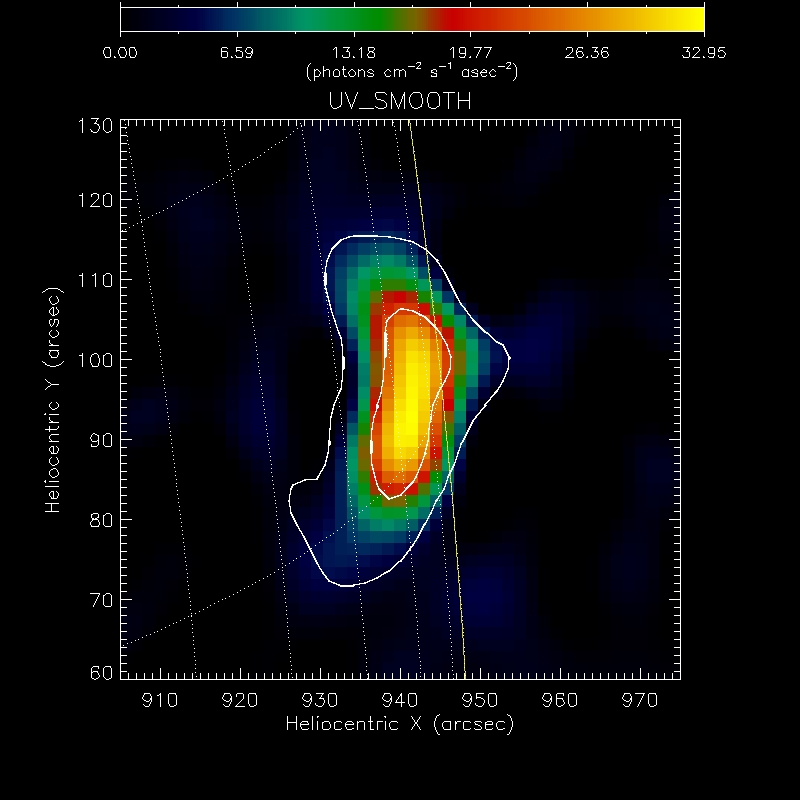}\\
                        EM & Clean & uv$\_$smooth
\end{tabular}
    \caption{Reconstructions of the August 31 2004 event provided by the same six imaging methods considered in Figures \ref{figure:fig3} through \ref{figure:fig5}. Time interval: 05:34:47 - 05:35:47; energy range: $6-12$ keV; detectors: $3$ through $9$. For sake of comparison, contour levels of MEM$\_$GE corresponding to $10\%$ and $50\%$ of the maximum intensity are superimposed to the reconstructions.}
    \label{figure:fig8}
\end{figure}

In order to illustrate the behavior of the new algorithm when applied to {\em{RHESSI}} observations, we consider two sets of events. The first set is made of the same cases considered in Figure \ref{figure:fig1} and we verified whether MEM$\_$GE is able to provide reconstructions that do not suffer the same pathological behavior characterizing MEM$\_$NJIT, while guaranteeing the same data fidelity. Specifically, Figure \ref{figure:fig2} refers to the same events considered in Figure \ref{figure:fig1}, but this time the reconstruction method employed is MEM$\_$GE. Then, Figures \ref{figure:fig3} through \ref{figure:fig5} compare the reconstructions provided by MEM$\_$GE and MEM$\_$NJIT with the ones of VIS$\_$CS, EM, Clean and uv$\_$smooth for these same three datasets. Finally, Figures \ref{figure:fig6} through \ref{figure:fig8} show the reconstructions provided by all six algorithms for three events whereby MEM$\_$NJIT works properly (for the reconstructions in Figures \ref{figure:fig3} through \ref{figure:fig8} we do not report the comparisons between the measured and predicted visibilities since they show very similar behaviors among the reconstruction methods, with the only exception of uv$\_$smooth, which is characterized by fitting performances systematically slightly worse). Tables \ref{table:tab1} and \ref{table:tab2} illustrate a quantitative comparison of performances from all codes and for all events considered in this sub-section.

\begin{table}[h]
\begin{tabular}{rcccc}
\toprule
				&reduced $\chi^2 (all)$		&reduced $\chi^2$ (used)			&flux					&time\\
\midrule
\midrule
\multicolumn{5}{c}{February 20 2002 ($20-50$ keV; $11:06:05 - 11:07:42$ UT)}\\
\midrule
MEM$\_$GE		&2.45					&2.99						&26.2				&22\\
MEM$\_$NJIT		&2.43					&2.54						&21.0				&11\\
VIS$\_$CS		&2.88					&3.52						&18.1				&3\\
CLEAN			&2.74					&3.30						&39.6				&7\\
EM				&2.39					&2.91						&18.4				&83\\
UV$\_$SMOOTH	&5.39					&7.21						&40.8				&1\\
\midrule
\midrule
\multicolumn{5}{c}{May 1 2002 ($3-6$ keV; $19:21:29 - 19:22:29$ UT)}\\
\midrule
MEM$\_$GE		&3.02					&3.93						&1.81 $\times 10^4$ 	&12\\
MEM$\_$NJIT		&2.90					&1.68						&1.49 $\times 10^4$		&12\\
VIS$\_$CS		&2.18					&2.64						&1.56 $\times 10^4$		&3\\
CLEAN			&2.30					&2.82						&2.95 $\times 10^4$		&11\\
EM				&1.88					&2.18						&1.56 $\times 10^4$		&100\\
UV$\_$SMOOTH	&3.19					&4.18						&1.94 $\times 10^4$		&1\\
\midrule
\midrule
\multicolumn{5}{c}{December 13 2007 ($6-12$ keV; $22:11:33 - 22:12:56$ UT)}\\
\midrule
MEM$\_$GE		&1.76					&1.61						&12.6 $\times 10^2$		&7\\
MEM$\_$NJIT		&7.45					&2.29						&5.63 $\times 10^2$		&59\\
VIS$\_$CS		&2.64					&3.01						&6.47 $\times 10^2$		&3\\
CLEAN			&2.68					&3.10						&9.90 $\times 10^2$		&8\\
EM				&2.56					&2.91						&6.57 $\times 10^2$		&108\\
UV$\_$SMOOTH	&2.75					&3.20						&8.01 $\times 10^2$		&1\\
\bottomrule
\end{tabular}
\caption{Quantitative parameters corresponding to the reconstructions of the three events presented in Figures \ref{figure:fig3} through \ref{figure:fig5}. The flux is measured in photon cm$^{-2}$ s$^{-1}$, time is measured in s.}
\label{table:tab1}
\end{table}

\begin{table}[h]
 \centering
\begin{tabular}{rcccc}
\toprule
				&reduced $\chi^2 (all)$		&reduced $\chi^2$ (used)			&flux					&time\\
\midrule
\midrule   \multicolumn{5}{c}{February 13 2002 ($6-12$ keV; $12:29:40 - 12:31:22$ UT)}\\
\midrule
MEM$\_$GE		&0.96					&0.99						&4.15 $\times 10^2$		&20\\
MEM$\_$NJIT		&0.98					&1.03						&4.22 $\times 10^2$		&3\\
VIS$\_$CS		&0.99					&1.04						&3.70 $\times 10^2$		&3\\
CLEAN			&1.07					&1.15						&6.67 $\times 10^2$		&8\\
EM				&0.97					&1.01						&3.67 $\times 10^2$		&45\\
UV$\_$SMOOTH	&2.13					&2.67						&7.38 $\times 10^2$		&1\\
\midrule
\midrule
\multicolumn{5}{c}{July 23 2002 ($25-50$ keV; $00:29:23 - 00:29:39$ UT)}\\
\midrule
MEM$\_$GE		&3.42					&3.53						&2.90 $\times 10^3$		 &20\\
MEM$\_$NJIT		&5.17					&6.18						&2.71 $\times 10^3$		&4\\
VIS$\_$CS		&5.45					&6.07						&2.41 $\times 10^3$		&3\\
CLEAN			&4.78					&5.04						&2.85 $\times 10^3$		&10\\
EM				&4.21					&4.67						&2.41 $\times 10^3$		&158\\
UV$\_$SMOOTH	&6.17					&6.86						&3.19 $\times 10^3$		&1\\
\midrule
\midrule
\multicolumn{5}{c}{August 31 2004 ($6-12$ keV; $05:34:47 - 05:35:47$ UT)}\\
\midrule
MEM$\_$GE		&1.24					&1.32						&1.44 $\times 10^4$		&14\\
MEM$\_$NJIT		&1.31					&1.42						&1.24 $\times 10^4$		&5\\
VIS$\_$CS		&1.20					&1.26						&1.29 $\times 10^4$		&3\\
CLEAN			&1.22					&1.29						&1.70 $\times 10^4$		&10\\
EM				&1.25					&1.33						&1.29 $\times 10^4$		&82\\
UV$\_$SMOOTH	&1.76					&2.08						&1.54 $\times 10^4$		&1\\
\bottomrule
\end{tabular}
\caption{Quantitative parameters corresponding to the reconstructions of the three events presented in Figures \ref{figure:fig6} through \ref{figure:fig8}. The flux is measured in photon cm$^{-2}$ s$^{-1}$, time is measured in s.}
\label{table:tab2}
\end{table}

%\begin{figure}[h]
 %   \includegraphics[width=0.4\textwidth]{mem_ge_1May2002_192129_3_6keV}
  %  \includegraphics[width=0.4\textwidth]{mem_njit_1May2002_192129_3_6keV}
   %\includegraphics[width=0.4\textwidth]{mem_ge_1May2002_192129_3_6keV}
    %\includegraphics[width=0.4\textwidth]{mem_njit_1May2002_192129_3_6keV}
    %\caption{Left: MEM\_GE, right: MEM\_NJIT}
%\end{figure}

\subsection{STIX}

\begin{figure}[h]
\centering
\includegraphics[height=3.5cm]{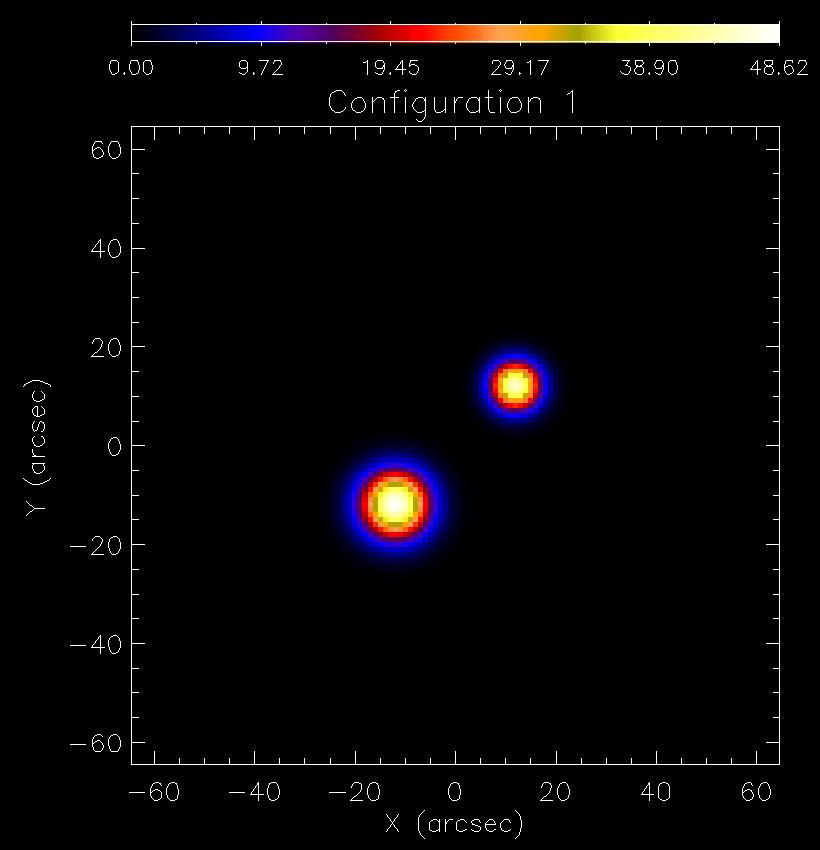}
\includegraphics[height=3.5cm]{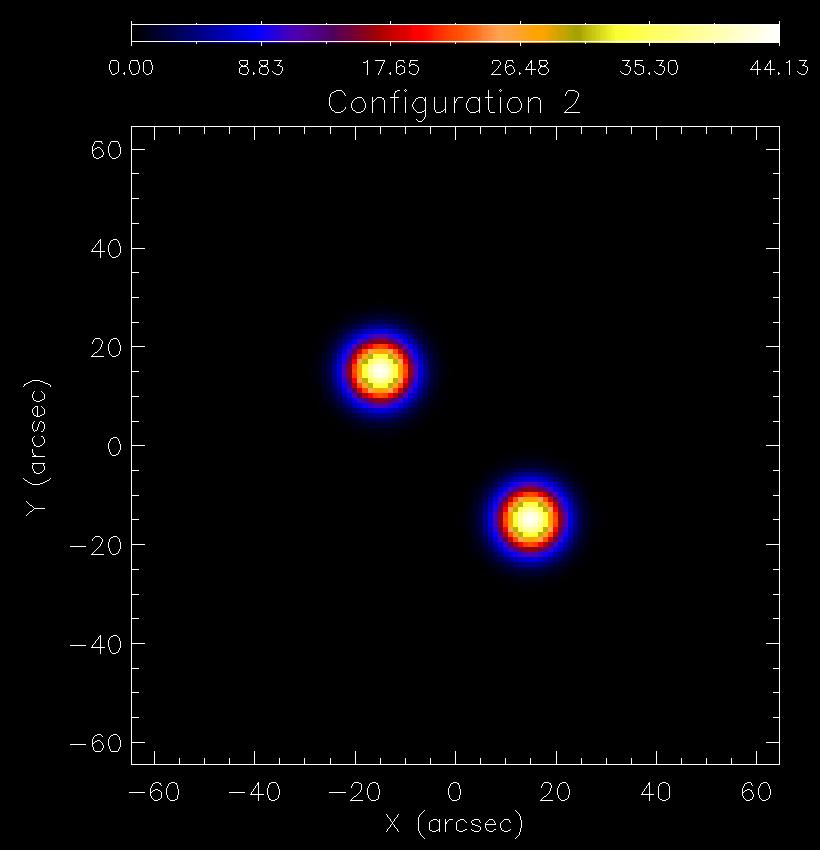}
\includegraphics[height=3.5cm]{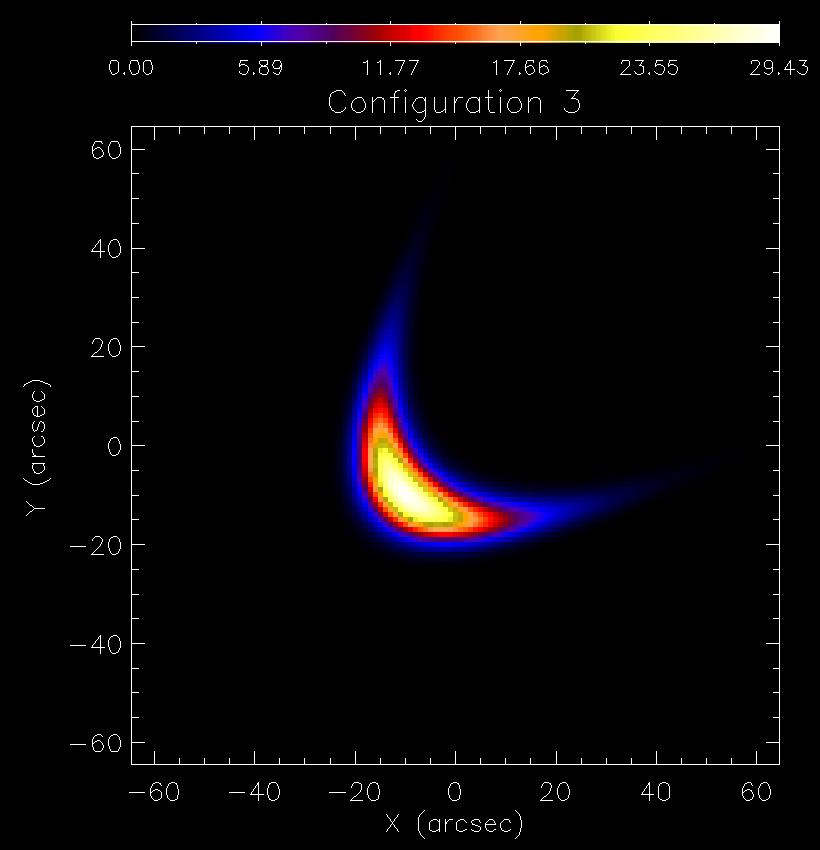}
\includegraphics[height=3.5cm]{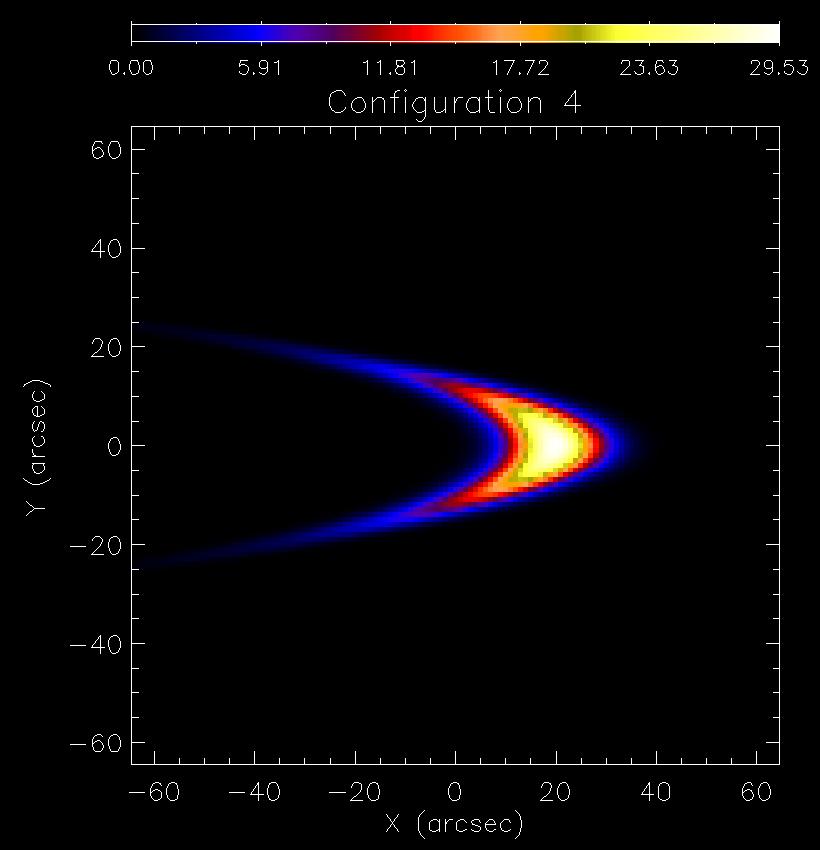}
\caption{Four ground-truth configurations utilized to generate synthetic {\em{STIX}} visibilities. Frome left to right: two foot-points with different size (Configuration 1); two foot-points with same size (Configuration2); a loop with orientation in the bottom left - top right direction (Configuration 3); a loop with orientation in the right - left direction (Configuration 4).}
\label{figure:fig9}
\end{figure}

We simulated the following four configurations with an overall incident flux of $10^4$ photons $\text{cm}^{-2}$ $\text{s}^{-1}$ (see Figure \ref{figure:fig9} and Table \ref{table:tab3} for all parameters):
\begin{itemize}
    \item a double footpoint flare in which one of the sources is more extended and has double the flux of the other source (Configuration 1);
    \item a double footpoint flare in which the sources have the same size and the same flux (Configuration 2);
    \item a loop flare with small curvature (Configuration 3);
    \item a loop flare with large curvature (Configuration 4).
\end{itemize}
For each one of these configurations the {\em{STIX}} software utilized a Monte Carlo approach to produce a set of synthetic visibilities with associated the corresponding standard deviations. These visibility sets have been processed by MEM$\_$GE and the results are compared in Table \ref{table:tab2} with the ones provided by the two other methods currently implemented in the {\em{STIX}} software tree, i.e. visibility-based Clean and count-based Expectation Maximization (EM) \citep{2019A&A...624A.130M}. 

\section{Discussion of results}
\begin{figure}
\begin{center}
\begin{tabular}{ccc}
\includegraphics[height=4.cm]{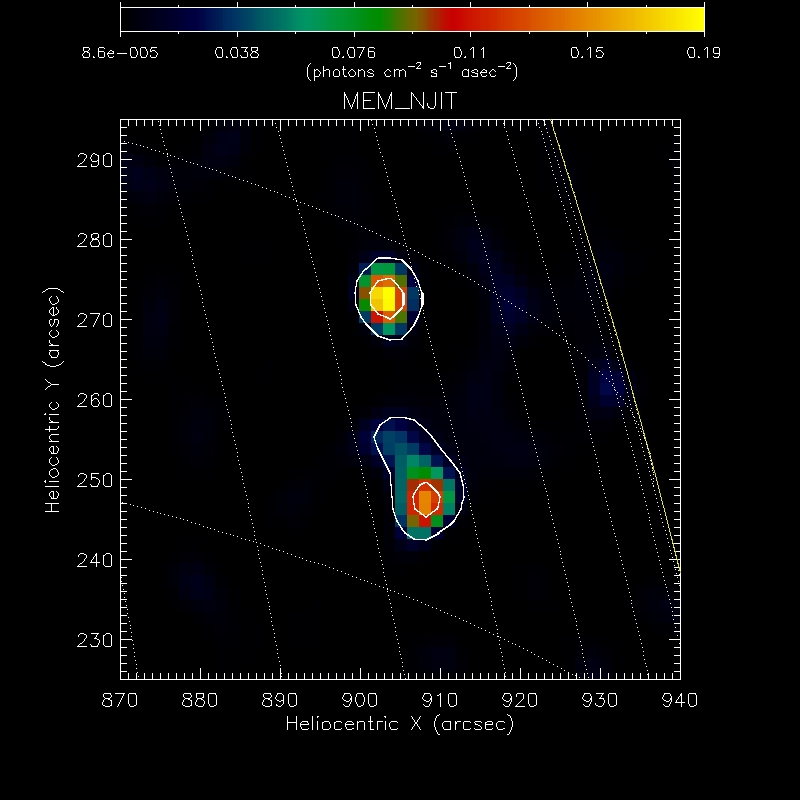} &
\includegraphics[height=4.cm]{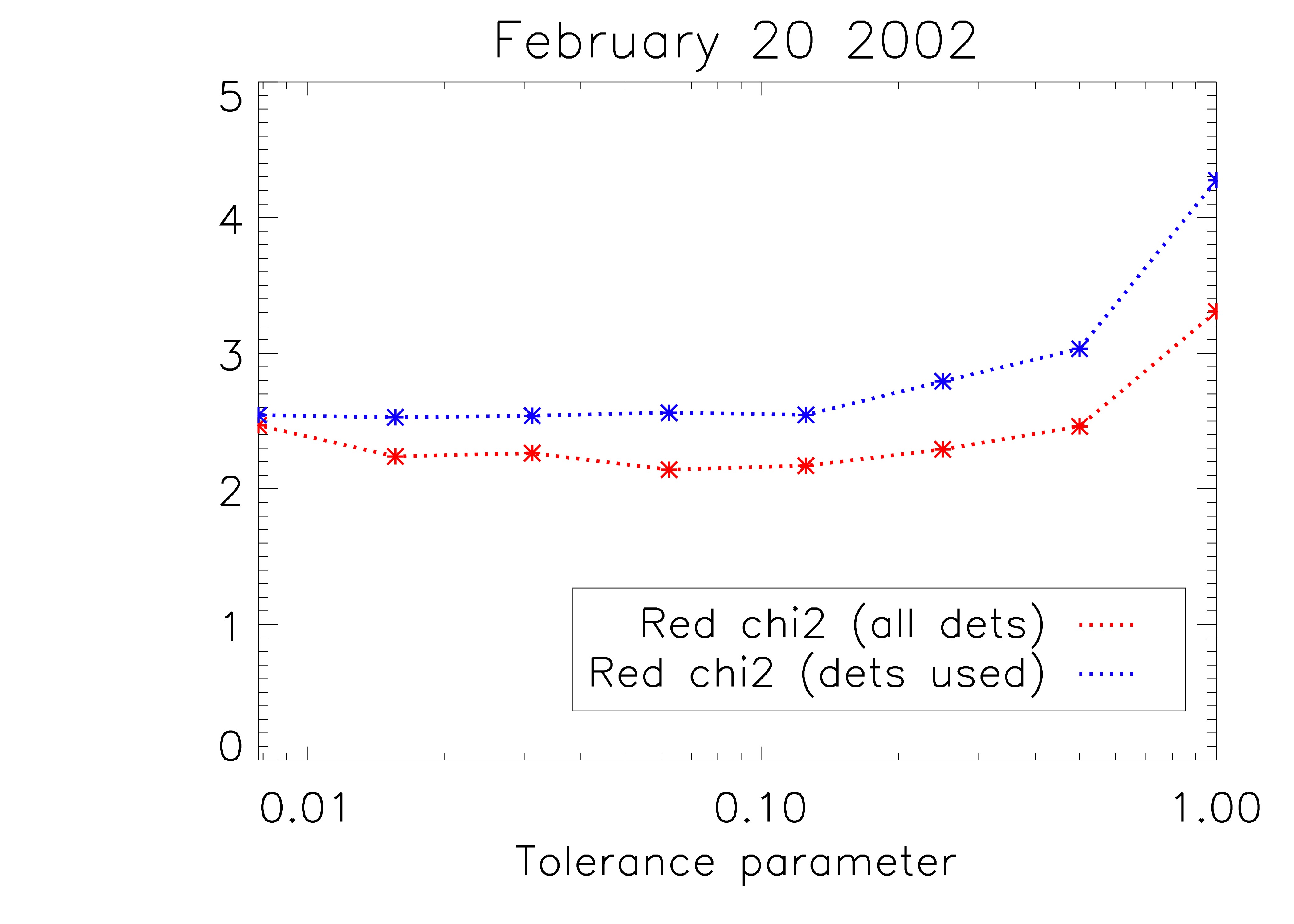} &
\includegraphics[height=4.cm]{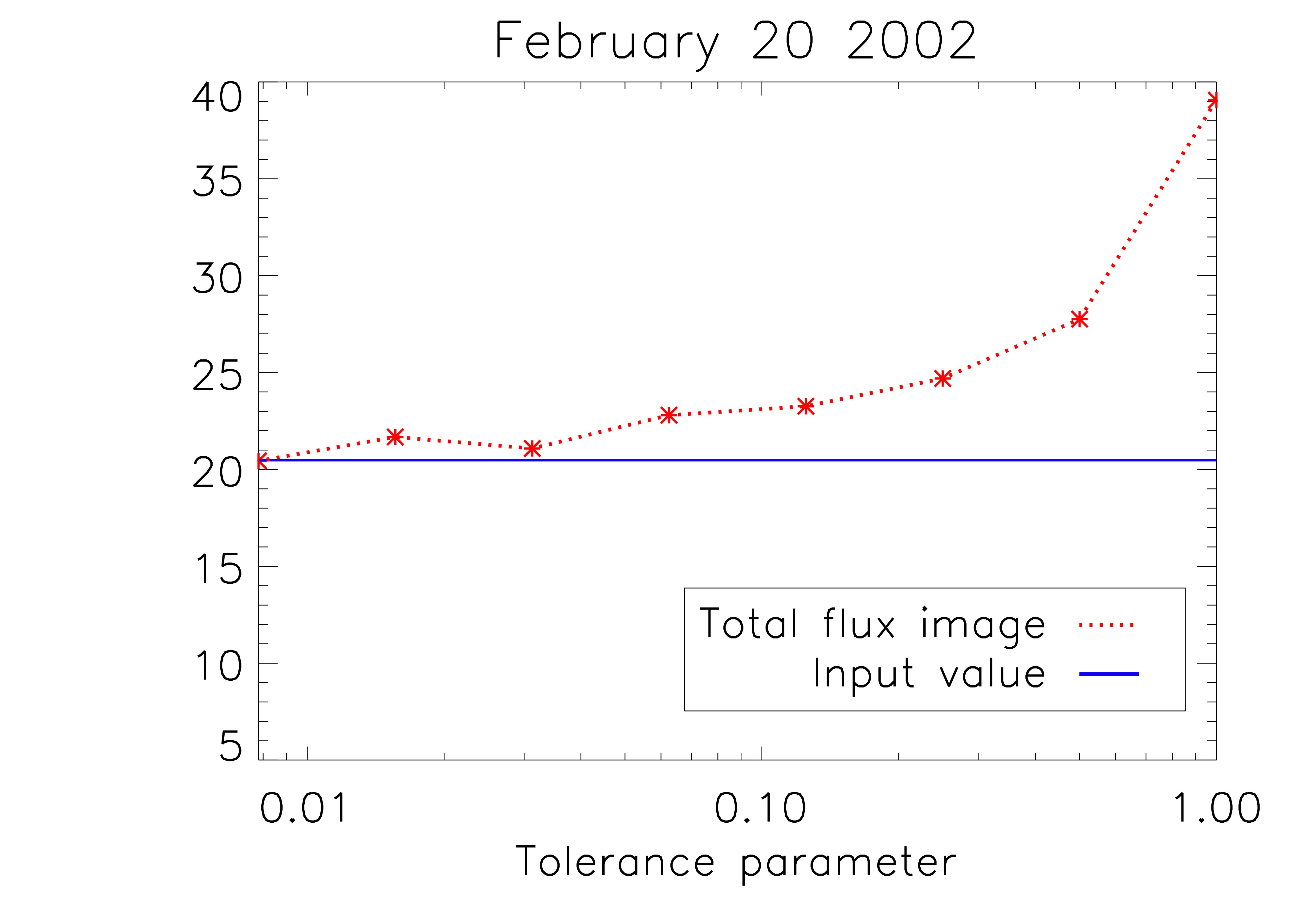} \\
\includegraphics[height=4.cm]{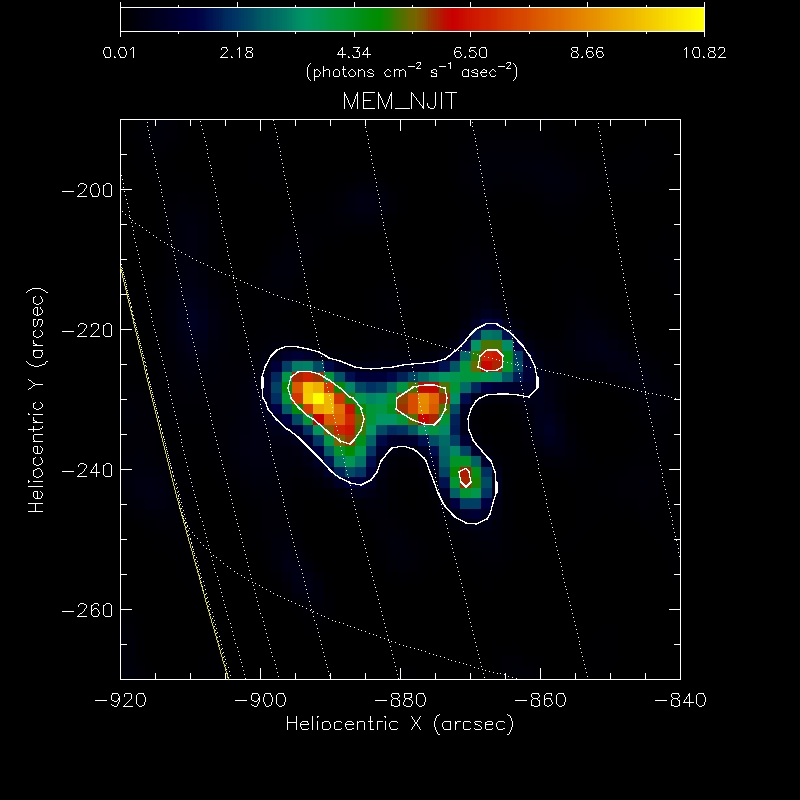} &
\includegraphics[height=4.cm]{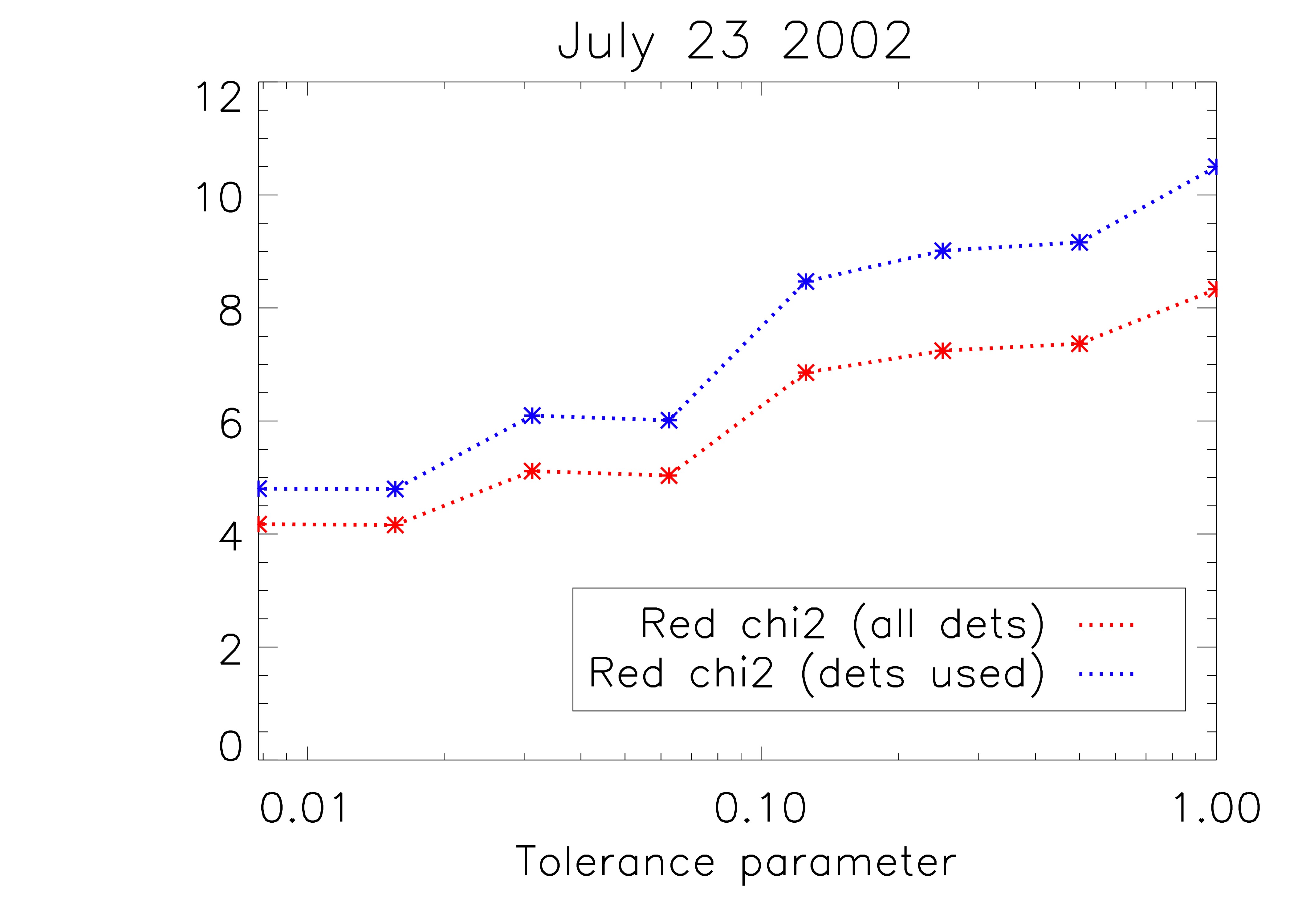} &
\includegraphics[height=4.cm]{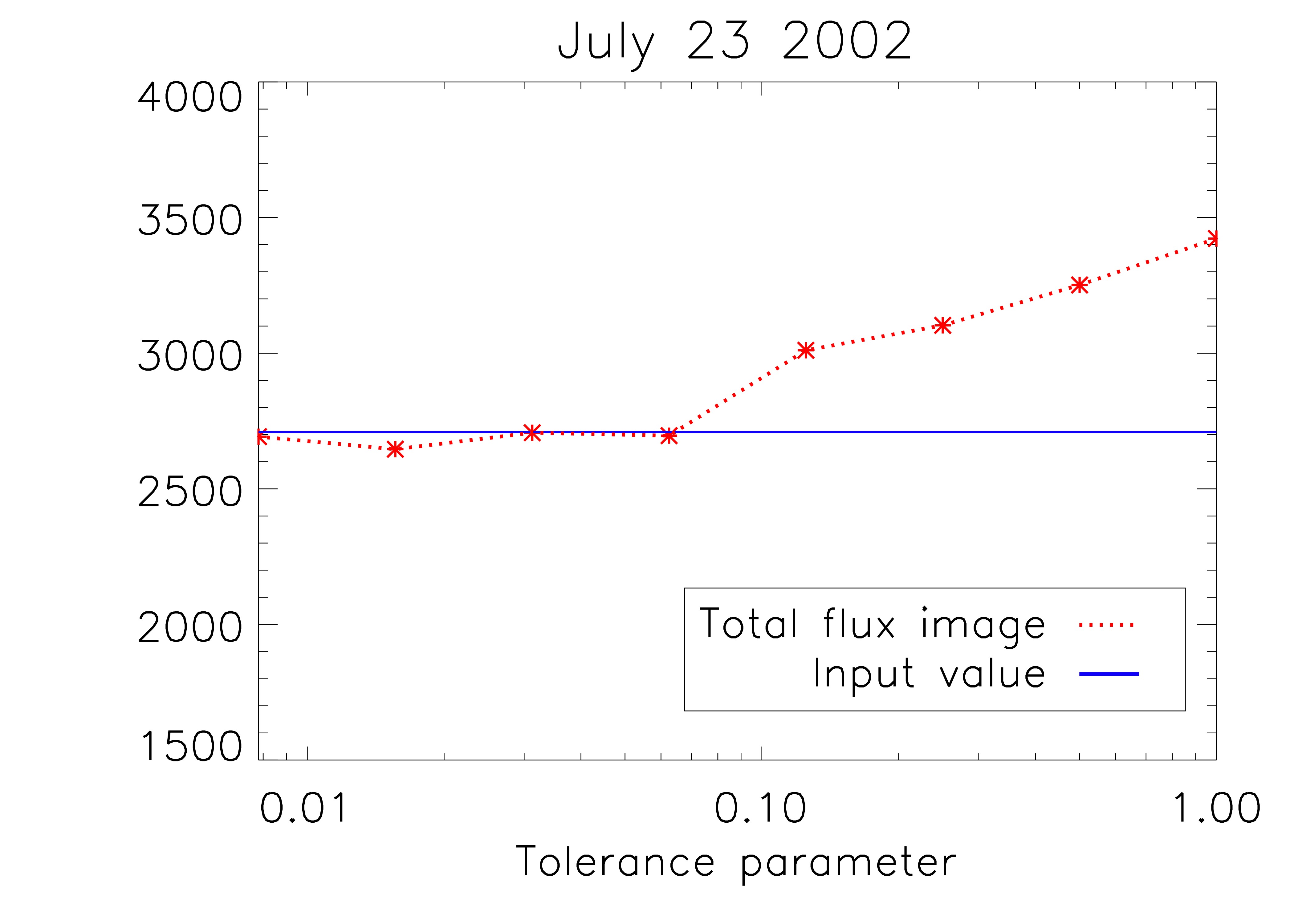}
\end{tabular}
\caption{Analysis of the outcome of MEM$\_$NJIT for different values of the tolerance parameter. First row: the February 20 2002 event. Second row: the July 23 2002 event. First column: MEM$\_$NJIT reconstructions for tolerance=$1$, with superimposed the level curves corresponding to MEM$\_$GE reconstructions; second column: reduced $\chi^2$ value vs tolerance computed considering both used and all detectors; third column: total flux of the source obtained from the image (red line) vs tolerance. The blue line corresponds to an a priori estimate used by MEM$\_$NJIT as input constraint.}
\label{figure:fig10}
\end{center}
\end{figure}

One of the nice aspects of MEM$\_$GE (see Figures \ref{figure:fig1} and \ref{figure:fig2}) is that it provides reliable reconstructions for those events where MEM$\_$NJIT, with its default tolerance parameter set to $0.03$, produces multiple unrealistically small sources, while predicting the experimental visibilities with a statistical fidelity close to the MEM$\_$NJIT one. On the other hand, it behaves similarly to MEM$\_$NJIT for those flaring events where MEM$\_$NJIT reconstructions are reliable (see Figures \ref{figure:fig6}, \ref{figure:fig7}, and \ref{figure:fig8}). As far as the morphological properties are concerned, MEM$\_$GE systematically introduces high-resolution effects. This is particularly visibile in the case of the reconstruction of the February 13 2002 event (Figure \ref{figure:fig6}), where a convex loop-shape is reproduced by MEM$\_$GE and MEM$\_$NJIT (and also EM), while this convexity is much smeared out in the reconstructions provided by Clean and uv$\_$smooth (however, we point out that the performances of Clean depend on what Regression Combined Method is used for the final Clean beam image and what Beam Width Factor is used). In this case, VIS$\_$CS fails to produce the correct orientation of the loop and reproduces a concave shape. Analogously, the spatial resolution achieved by the two MEM codes (and by EM and, partly, by uv$\_$smooth) is as fine as can be expected in the reconstruction of the July 23 2002 X Class flare, given the angular resolution of the modulation collimators used in the analysis: all four sources characterizing this emission topography are clearly visibile in the reconstructions. On the contrary, neither VIS$\_$CS nor Clean are able to distinguish all four emitting regions. Tables \ref{table:tab1} and \ref{table:tab2} show that, when MEM$\_$NJIT works properly, the two maximum entropy methods provide similar estimates of the quantitative parameters, although MEM$\_$GE may reproduce the data with significantly smaller $\chi^2$ values (as in the case of the December 13 2007 and July 23 2002 events) but may require a higher computational time (as in the case of the February 13 2002 and July 23 2002 events). We finally point out that the tolerance parameter in MEM$\_$NJIT can be manually tuned to higher values in order to obtain more regularized reconstructions. However, Figure \ref{figure:fig10} shows that increasing the tolerance value results into morphologies closer to the ones corresponding to MEM$\_$GE reconstructions but with a less accurate ability of the method to both fit the observations and conserve the flux value given as input to the algorithm. Even more importantly, the figure shows that this tuning procedure is dependent on the experimental dataset considered.

Interestingly, the super-resolution properties of Maximum Entropy are confirmed by the analysis of synthetic {\em{STIX}} data illustrated in Table \ref{table:tab3}, where the performances of MEM$\_$GE are compared to the ones of visibility-based Clean and count-based EM (there is no MEM$\_$NJIT code adapted to the {\em{STIX}} framework). In fact, in the reconstructions of the foot-points, the ability of MEM$\_$GE to recover the ground-truth full width at half maximum (FWHM) outperforms the effectiveness of the other visibility-based algorithm (EM works systematically better, probably because the signal-to-noise ratio associated to counts is higher than the ones associated to the real and imaginary parts of the visibilities, as shown by \citet{2019A&A...624A.130M}).

Also, similarly to EM, MEM$\_$GE behaves better than Clean in both reproducing the exact total flux and best-fitting the synthetic measurements. EM and Clean seem more accurate in separately reproducing the flux of each one of the two circular sources in Configuration 1 and Configuration 2.

\section{Conclusions}
This paper introduces a novel algorithm, named MEM$\_$GE, implementing a Maximum Entropy approach to image reconstruction from X-ray visibilities in solar astronomy. The motivation of this effort relies on the fact that the only existing algorithm (MEM$\_$NJIT) realizing this approach for this kind of data provides reliable and super-resolved reconstructions for datasets associated to most events, but sometimes suffers numerical instabilities and lack of convergence to meaningful solutions. Differently than the old algorithm, the new implementation relies on the constrained minimization of a convex functional, which is realized by alternating gradient descent and proximal steps. On the one hand, the resulting code maintains the good imaging properties of the previous one, while guaranteeing, on the other hand, convergence to reliable reconstructions when MEM$\_$NJIT gives unphysical sources (unless the tolerance parameter is tuned in a data-dependent fashion).

MEM$\_$GE is available in the Solar SoftWare tree and can be most easily accessed through the HESSI GUI. It is one of the algorithms that is currently used for the population of the {\em{RHESSI}} image archive, and can be used also for the processing of {\em{STIX}} synthetic visibilities.

The nice imaging properties of this algorithm, together with its systematic reliability, make it particularly appropriate for the accurate estimation of morphological properties like the ones considered in the analysis of coronal hard X-ray sources \citep{2008ApJ...673..576X,2012ApJ...755...32G,2012A&A...543A..53G,2018ApJ...867...82D}. MEM$\_$GE is an appropriate algorithm to realize a statistical study of these kinds of events.

\begin{table}[ht]
\centering	
\resizebox{\linewidth}{!}{\begin{tabular}{rcccccccccc} 

\toprule

\multicolumn{11}{c}{Configuration 1}\\

\midrule
	 			&\multicolumn{4}{c}{First Peak}												&\multicolumn{4}{c}{Second Peak} 		&Total flux ($\times 10^3$)	&C-statistic\\							
\cmidrule(lr){2-5}
\cmidrule(lr){6-9}
	 			
	 			&X				&Y		  		&FWHM			&Flux ($\times 10^3$)	&X			   	&Y		 		&FWHM			&Flux ($\times 10^3$)	&	&\\	
\cmidrule{2-11}
Simulated			&$-12.0$			&$-12.0$		  	&$11.0$			&$6.53$				&$12.0$			&$12.0$		    	&$8.0$			&$3.33$				&$10.00$					&\\
MEM$\_$GE		&$-12.0\pm0.5$ 	&$-11.8\pm0.5$ 	&$10.9\pm1.1$		&$4.79\pm0.12$		&$12.0\pm0.6$		&$11.7\pm0.6$		&$8.7\pm0.9$		&$2.24\pm0.11$		&$10.69\pm0.30$			&$6.0\pm1.2$\\
EM				&$-12.1\pm0.6$	&$-12.0\pm0.5$	&$11.1\pm0.9$		&$5.96\pm0.13$		&$12.0\pm0.5$		&$12.0\pm0.5$		&$8.3\pm0.4$		&$2.95\pm0.10$		&$10.63\pm0.04$			&$3.4\pm0.3$\\
CLEAN		&$-12.3\pm0.6$	&$-12.0\pm0.4$	&$13.5\pm0.5$		&$5.34\pm0.10$		&$11.8\pm0.5$		&$12.1\pm0.3$		&$10.1\pm0.2$		&$2.59\pm0.09$		&$8.59\pm0.12$			&$26.5\pm2.9$\\

\midrule
\multicolumn{11}{c}{Configuration 2}\\

\midrule

	 			&\multicolumn{4}{c}{First Peak}													&\multicolumn{4}{c}{Second Peak} 		&Total flux ($\times 10^3$)	&C-statistic\\
\cmidrule(lr){2-5}
\cmidrule(lr){6-9}
	 			
	 			&X				&Y		  		&FWHM			&Flux ($\times 10^3$)		&X			   	&Y		 		&FWHM			&Flux ($\times 10^3$)	&	&\\
\cmidrule{2-11}
Simulated			&$-15.0$			&$15.0$		  	&$10.0$			&$4.95$					&$15.0$			&$-15.0$		    	&$10.0$			&$4.95$				&$10.00$		&\\
MEM$\_$GE		&$-14.6\pm0.7$ 	&$14.7\pm0.5$ 	&$8.3\pm0.5$		&$3.37\pm0.13$			&$14.5\pm0.7$		&$-14.8\pm0.4$	&$8.2\pm0.5$		&$3.41\pm0.12$		&$10.64\pm0.21$		&$5.9\pm0.8$\\
EM				&$-14.4\pm0.6$	&$14.8\pm0.5$		&$8.7\pm0.6$		&$4.30\pm0.14$			&$14.4\pm0.6$		&$-14.8\pm0.6$	&$8.5\pm0.6$		&$4.32\pm0.13$ 		&$10.61\pm0.03$		&$3.2\pm0.3$\\
CLEAN		&$-13.5\pm0.5$	&$14.6\pm0.5$		&$12.1\pm0.4$		&$3.82\pm0.10$			&$13.4\pm0.5$		&$-14.4\pm0.6$	&$11.9\pm0.4$		&$3.84\pm0.10$		&$8.38\pm0.11$		&$31.2\pm2.5$\\
\midrule
	 			
\multicolumn{11}{c}{Configuration 3}\\
	 			
\midrule
	 			&\multicolumn{2}{c}{Peak}				&Total flux ($\times 10^3$)			&C-statistic  		&	&	&	&	&	&\\
\cmidrule(lr){2-3}
	 			&X				&Y				&								&				&	&	&	&	&	&\\
\cmidrule{2-6}
Simulated			&$-10.0$			&$-10.0$			&10.00							&			 	&	&	&	&	&	&\\
MEM$\_$GE		&$-9.1\pm1.8$		&$-10.0\pm2.1$	&$10.55\pm0.21$					&$6.1\pm0.6$		&	&	&	&	&	&\\
EM				&$-9.6\pm2.0$		&$-9.8\pm2.7$		&$10.64\pm0.04$					&$3.4\pm0.3$		& 	&	&	&	&	&\\
CLEAN		&$-10.2\pm1.3$	&$-9.0\pm2.0$		&$8.51\pm0.11$					&$29.0\pm2.6$		& 	&	&	&	&	&\\
				
\midrule
	 			
\multicolumn{11}{c}{Configuration 4}\\
	 			
\midrule
	 			&\multicolumn{2}{c}{Peak}				&Total flux ($\times 10^3$)			&C-statistic  		&	&	&	& 	&	&\\
\cmidrule(lr){2-3}
	 			&X				&Y				&								&				&	&	&	&	&	&\\
\cmidrule{2-6}
Simulated			&$20.0$			&$0.0$			&10.00							&			 	&	&	&	&	&	&\\
MEM$\_$GE		&$18.2\pm1.0$		&$0.2\pm0.7$		&$10.44\pm0.17$					&$5.8\pm0.3$		&	&	&	&	&	&\\
EM				&$18.5\pm1.1$		&$0.4\pm0.7$		&$10.63\pm0.03$					&$3.4\pm0.2$		& 	&	&	&	&	&\\
CLEAN		&$18.6\pm0.7$		&$0.0\pm0.3$		&$8.39\pm0.11$					&$32.1\pm2.9$		& 	&	&	&	&	&\\		
\bottomrule
\end{tabular}}
\caption{Reconstruction of four source configurations characterized by an overall incident photon flux of $10^4$ photons $\text{cm}^{-2}$ $\text{s}^{-1}$. The morphological and photometric parameters reconstructed by MEM$\_$GE are compared with the ground truth and with the values provided by EM and CLEAN. The positions X, Y and the full width at half maximum (FWHM) of the sources are given in arcseconds, while the total flux and the flux corresponding to each foot point in Configurations 1 and 2 are given in photons $\text{cm}^{-2}$ $\text{s}^{-1}$. Mean values and standard deviations are computed on 25 random realizations of the data performed with the simulator implemented in the STIX DPS.}
\label{table:tab3}
\end{table}

\bibliography{biblio.bib}
\bibliographystyle{aasjournal}

\end{document}